\documentclass[11pt]{article}
\usepackage{epsfig} 
\newcommand{\sect}[1]{ \section{#1} \setcounter{equation}{0} }

\newcommand{\Dslash}{D \! \! \! \! /} 
 
\newcommand{\pslash}{p \! \! \! /}

\newcommand{\half}{\mbox{\small{$\frac{1}{2}$}}} 
 
\newcommand{\Nf}{N_{\!f}} 
\newcommand{\MSbar}{\overline{\mbox{MS}}} 
\newcommand{\MOMbar}{\overline{\mbox{MOM}}} 

\begin{document}
\title{Three loop anomalous dimension of non-singlet quark currents in the
RI${}^\prime$ scheme} 
\author{J.A. Gracey, \\ Theoretical Physics Division, \\ 
Department of Mathematical Sciences, \\ University of Liverpool, \\ P.O. Box 
147, \\ Liverpool, \\ L69 3BX, \\ United Kingdom.} 
\date{} 
\maketitle 
\vspace{5cm} 
\noindent 
{\bf Abstract.} We renormalize QCD at three loops in the modified 
regularization invariant, RI$^\prime$, scheme in arbitrary covariant gauge and
deduce that the four loop $\beta$-function is equivalent to the $\MSbar$
result. The anomalous dimensions of the scalar, vector and tensor currents are 
then determined in the RI$^\prime$ scheme at three loops by considering the 
insertion of the operator in a quark two-point function. The expression for the
scalar current agrees with the quark mass anomalous dimension and we deduce an
expression for the four loop RI$^\prime$ mass anomalous dimension in arbitrary
covariant gauge and for any Lie group. 

\vspace{-18cm}
\hspace{13.5cm}
{\bf LTH 572}

\newpage

\sect{Introduction.} 

The most widely used renormalization prescription in perturbative quantum field
theory is the minimal subtraction scheme where only the infinities with respect
to the regularization are subtracted from the divergent part of the Green's 
function to determine the renormalization group functions, \cite{1}. In 
practice, though, it is more appropriate to use the modified minimal 
subtraction scheme, $\MSbar$, since the convergence properties of perturbative 
series in this scheme are improved by additionally absorbing a {\em finite} 
part, $\ln(4\pi e^{-\gamma})$, into the renormalization constants, \cite{2}, 
where $\gamma$ is the Euler-Mascheroni constant. The advantage of using the 
$\MSbar$ scheme, which is a mass independent scheme, rests in some elegant 
properties. For example, the $\MSbar$ $\beta$-function and anomalous dimension 
of the quark mass in QCD, are both independent of the covariant gauge fixing
parameter, \cite{1,3}. Moreover, performing computations using $\MSbar$ and 
dimensional regularization, where the spacetime dimension becomes 
$d$~$=$~$4$~$-$~$2\epsilon$ and $\epsilon$ is the regularizing parameter, one 
can carry out multiloop calculations to very high order. Indeed in QCD various 
four loop renormalization group functions are available, \cite{4,5,6}, which 
represent the current state of computation. Whilst the $\MSbar$ scheme enjoys 
these elegant features and has become the standard reference scheme, it has the
limitation that it is not a physical renormalization scheme. Examples of 
schemes which are founded in a more physical origin include, for example, the 
MOM and $\MOMbar$ schemes, \cite{7,8}. Though one disadvantage of using 
physical schemes is that fewer results currently exist to the same multiloop 
precision as in $\MSbar$. However, it is well known that physical quantities in
one scheme can be simply related to the same quantity in other schemes by a 
conversion function, \cite{7}. Whilst one ordinarily uses dimensionally 
regularized perturbation theory to compute physical quantities one can also 
determine such information by using a lattice regularization. The advantage of 
this approach is that one in principle includes all non-perturbative 
contributions in a calculation which need to be converted from the lattice
scheme to $\MSbar$. For a recent review and applications in determining matrix 
elements in deep inelastic scattering see, for example, \cite{9} where lattice 
results were matched to $\MSbar$ results. The scheme used is similar to a 
modified version of the regularization invariant, RI, scheme known as the 
RI${}^\prime$ scheme, \cite{10}. Therefore, in order to improve lattice 
estimates one requires the conversion of various renormalization group 
functions from $\MSbar$ to RI$^\prime$. Recently, this problem has been 
addressed in the context of quark masses where the conversion functions were 
produced for all covariant gauges for the quark mass anomalous dimension at 
four loops, \cite{11,12}. Indeed the field anomalous dimensions were also 
deduced to the same order for the $SU(N_{\! c})$ colour group, \cite{12}. 
Though in practice for the lattice application one only considers one 
particular gauge which is the Landau gauge. This work of \cite{12} extended the
three loop calculation of \cite{11}. One practical feature of these 
computations was that one only needed to consider ordinary perturbation theory 
in the massless limit which effectively meant that the conversion functions 
could be deduced using standard multiloop perturbative tools for massless field
theories. 

Whilst these papers dealt with the problem of quark masses deduced from the
lattice there are other problems where the conversion functions are required. 
For instance, there is interest in deducing low moments of the structure 
functions measured in deep inelastic scattering from the lattice, \cite{9,13}. 
To improve estimates the conversion factors from the $\MSbar$ scheme to the 
RI$^\prime$ scheme are required. Therefore, the purpose of this article is 
twofold. First, given that the RI$^\prime$ scheme is important for relating 
Landau gauge lattice results to the $\MSbar$ scheme we will renormalize QCD at 
three loops in the RI$^\prime$ scheme though in a general covariant gauge. 
Whilst we are ultimately interested in quark currents it is not inconceivable 
that the anomalous dimensions of operators with gluonic fields will at some 
time be measured on the lattice and therefore the anomalous dimensions of the 
gluon (and ghost) fields will need to be determined at the same level as the 
quarks. Equipped with the fully renormalized QCD Lagrangian in the RI$^\prime$ 
scheme we will then extend the approach of \cite{12} to deduce RI$^\prime$ 
information but for the anomalous dimension of a particular quark composite 
operator which corresponds to the tensor current in QCD. It is of interest 
since it represents the lowest moment of the transversity operator in deep 
inelastic scattering, \cite{14}. Given the recent resurgence of experimental 
and theoretical interest in transversity, (see, for example, 
\cite{15,16,17,18,19}), the longer term aim is to provide a more accurate 
numerical estimate for the associated matrix element prior to experimental data
being accumulated at RHIC. One motivation in this approach is to develop the 
calculational formalism to determine the conversion function for an operator 
which is a simple extension of \cite{12}. In \cite{12} the quark mass 
conversion functions were determined by considering the corrections to the 
massive quark two-point function which is an avenue not immediately available 
for a composite operator. Nevertheless we will bridge this gap by 
reconstructing the result of \cite{12} at three loops by first considering the 
problem of the mass renormalization as the renormalization of the associated 
composite operator, $\bar{\psi} \psi$, as inserted in a two-point Green's 
function prior to replacing it by the operator of main interest which is the 
tensor current. Moreover, given this way of computing we need only use the 
{\em massless} version of QCD. 

The paper is organised as follows. In section two we discuss the three loop
renormalization of QCD in the RI$^\prime$ scheme and provide the 
renormalization group functions. These results are used in section three to 
extract the anomalous dimension of the quark mass operator in QCD in 
RI$^\prime$ which agrees with the earlier Landau gauge result of \cite{12}. 
Having provided the formalism for treating an operator, we extend that 
calculation to the tensor current case in section four. Finally, our 
conclusions are given in section five. 

\sect{RI$^\prime$ scheme at three loops.} 

We begin by explicitly renormalizing QCD in the RI$^\prime$ scheme at three
loops using dimensional regularization. The bare QCD Lagrangian, with the gauge
fixed covariantly, is 
\begin{equation} 
L ~=~ -~ \frac{1}{4} G_{\mu\nu}^a G^{a \, \mu\nu} ~-~ \frac{1}{2\alpha} 
(\partial^\mu A^a_\mu)^2 ~-~ \bar{c}^a \partial^\mu D_\mu c^a ~+~ 
i \bar{\psi}^{iI} \Dslash \psi^{iI} 
\label{qcdlag}
\end{equation} 
where $A^a_\mu$ is the gluon field, $G^a_{\mu\nu}$ $=$ $\partial_\mu A^a_\nu$ 
$-$ $\partial_\nu A^a_\mu$ $-$ $g f^{abc} A^b_\mu A^c_\nu$, $c^a$ and 
$\bar{c}^a$ are respectively the ghost and antighost fields and $\alpha$ is the
covariant gauge fixing parameter. The indices range over 
$1$~$\leq$~$a$~$\leq$~$N_A$, $1$~$\leq$~$I$~$\leq$~$N_F$ and 
$1$~$\leq$~$i$~$\leq$~$\Nf$ where $N_F$ and $N_A$ are the respective dimensions
of the fundamental and adjoint representations of the colour group whose 
generators are $T^a$ and structure functions are $f^{abc}$ whilst $\Nf$ is the 
number of quark flavours. The covariant derivatives are defined by 
\begin{eqnarray} 
D_\mu \psi &=& \partial_\mu \psi ~+~ ig A^a_\mu T^a \psi ~~~,~~~  
D_\mu A^a_\nu ~=~ \partial_\mu A^a_\nu ~-~ g f^{abc} A^b_\mu A^c_\nu 
\nonumber \\ 
D_\mu c^a &=& \partial_\mu c^a ~-~ g f^{abc} A^b_\mu c^c ~.  
\end{eqnarray} 
In renormalizing the full Lagrangian in the RI$^\prime$ scheme, which we will
define more precisely later, we must ensure that the scheme is consistent. We
can achieve this, for example, by demonstrating that the renormalization 
constants for the gluon, ghost and quark fields correctly produce the 
{\em same} gauge independent coupling constant renormalization at three loops 
from different Green's functions, thereby ensuring that the Slavnov-Taylor 
identities are respected. If we regard all the quantities in the QCD 
Lagrangian, (\ref{qcdlag}), as bare and denote them with the subscript 
${}_{\mbox{\footnotesize{o}}}$, we introduce the corresponding renormalized 
quantities by the usual definitions 
\begin{equation} 
A^{a \, \mu}_{\mbox{\footnotesize{o}}} ~=~ \sqrt{Z_A} \, A^{a \, \mu} ~~,~~ 
c^a_{\mbox{\footnotesize{o}}} ~=~ \sqrt{Z_c} \, c^a ~~,~~ 
\psi_{\mbox{\footnotesize{o}}} ~=~ \sqrt{Z_\psi} \psi ~~,~~  
g_{\mbox{\footnotesize{o}}} ~=~ \mu^\epsilon Z_g \, g ~~,~~ 
\alpha_{\mbox{\footnotesize{o}}} ~=~ Z^{-1}_\alpha Z_A \, \alpha 
\label{Zdef}
\end{equation} 
where $\mu$ is the mass scale introduced to ensure the coupling constant is 
dimensionless in $d$-dimensions and $d$ $=$ $4$ $-$ $2\epsilon$ with $\epsilon$
the regularizing parameter.  

To determine the RI$^\prime$ scheme values of the renormalization constants we
first consider the gluon, quark and ghost two-point functions. In 
\cite{10,11,12} the RI$^\prime$ scheme definition of the quark wave function 
renormalization is given by the Minkowski space condition 
\begin{equation} 
\left. \lim_{\epsilon \rightarrow 0} \left[ 
Z^{\mbox{\footnotesize{RI$^\prime$}}}_\psi \Sigma_\psi(p) \right] 
\right|_{p^2 \, = \, \mu^2} ~=~ \pslash 
\label{Zpsidef}
\end{equation}  
where $\Sigma_\psi(p)$ is the bare (massless) quark two-point function and $p$ 
is the external quark momentum. As we are considering massless quarks 
$\Sigma_\psi(p)$ will be proportional to $\pslash$ and involve poles in 
$\epsilon$ at each order in the strong coupling constant. To contrast with the 
$\MSbar$ scheme, the RI$^\prime$ scheme definition of the quark wave function 
is such that one absorbs the complete finite part of the Green's function with 
respect to $\epsilon$ into the renormalization constant. In other words only
the $O(1)$ piece is removed and the $O(\epsilon)$ part is ignored. In the 
$\MSbar$ scheme only the poles in $\epsilon$ are removed as well as the finite 
parts involving powers of $\ln(4\pi e^{-\gamma})$. For completeness we note 
that in the RI scheme one absorbs the full finite part in the same way as in 
the RI$^\prime$ scheme, (\ref{Zpsidef}), but for a different part of the 
Green's function which is,  
\begin{equation} 
\left. \lim_{\epsilon \rightarrow 0} \left[ \frac{1}{4d} \mbox{tr} \left( 
Z^{\mbox{\footnotesize{RI}}}_\psi \gamma^\mu \frac{\partial ~}{\partial p^\mu} 
\Sigma_\psi(p) \right) \right] \right|_{p^2 \, = \, \mu^2} ~=~ 1 ~.  
\end{equation}  
Due to the presence of the derivative this scheme is much more difficult to 
implement on the lattice compared with (\ref{Zpsidef}) which is why we are
concentrating on RI$^\prime$. For the remaining wave function renormalization 
constants we define their RI$^\prime$ values in a similar way to 
(\ref{Zpsidef}). For the ghost fields $Z^{\mbox{\footnotesize{RI$^\prime$}}}_c$
is determined by  
\begin{equation} 
\left. \lim_{\epsilon \rightarrow 0} \left[ 
Z^{\mbox{\footnotesize{RI$^\prime$}}}_c \frac{\Sigma_c(p)}{p^2} \right] 
\right|_{p^2 \, = \, \mu^2} ~=~ 1 ~.  
\end{equation} 
The definition of the gluon field requires more care due to it having 
transverse and longitudinal components. Rendering the former finite determines
$Z_A$ whilst the finiteness of the latter component fixes the gauge parameter
renormalization constant. In the spirit of the quark and ghost RI$^\prime$
scheme renormalization constants we now define 
$Z^{\mbox{\footnotesize{RI$^\prime$}}}_A$ and
$Z^{\mbox{\footnotesize{RI$^\prime$}}}_\alpha$ through the following 
conditions. Writing the gluon polarization tensor, $\Pi_{\mu\nu}(p)$, as
\begin{equation}
\Pi_{\mu\nu}(p) ~=~ \frac{\Pi_T(p)}{p^2} \left[ \eta_{\mu\nu} ~-~ 
\frac{p_\mu p_\nu}{p^2} \right] ~+~ \Pi_L(p) \frac{p_\mu p_\nu}{(p^2)^2} 
\end{equation} 
we define the associated RI$^\prime$ scheme renormalization constants as
\begin{equation}
\left. \lim_{\epsilon \rightarrow 0} \left[ 
Z^{\mbox{\footnotesize{RI$^\prime$}}}_A \Pi_T(p) \right] 
\right|_{p^2 \, = \, \mu^2} ~=~ 1
\end{equation} 
and  
\begin{equation}
\left. \lim_{\epsilon \rightarrow 0} \left[ 
Z^{\mbox{\footnotesize{RI$^\prime$}}}_\alpha \Pi_L(p) \right] 
\right|_{p^2 \, = \, \mu^2} ~=~ 1 ~. 
\end{equation} 

Having established the scheme definition of the field renormalization constants
we have computed them explicitly for arbitrary $\alpha$ for (\ref{qcdlag}). We
have used the {\sc Mincer} package, \cite{20}, written in the symbolic 
manipulation language {\sc Form}, \cite{21,22}, where the Feynman diagrams are 
generated with {\sc Qgraf}, \cite{23}. The basic nature of the renormalization 
conditions are straightforward to implement in {\sc Form} using the approach of
\cite{24}. Briefly, one computes the appropriate Green's functions in terms of
bare parameters and couplings and then introduces the counterterms by rescaling
these variables into the renormalized ones using the definition of
renormalization constants, (\ref{Zdef}). Also, for Green's functions which 
ordinarily involve a bare parameter in the tree part, one divides the Green's
function by the bare parameter first before rescaling, \cite{24}. Therefore, we
quote the results of our calculations. We find  
\begin{eqnarray} 
Z^{\mbox{\footnotesize{RI$^\prime$}}}_A &=& 1 ~+~ 
\left[ \left( \left( \frac{13}{6} - \frac{\alpha}{2} \right) C_A 
- \frac{4}{3} T_F \Nf \right) \frac{1}{\epsilon} + \left( \left( \frac{97}{36}
+ \frac{\alpha}{2} + \frac{\alpha^2}{4} \right) C_A - \frac{20}{9} T_F\Nf 
\right) \right] a \nonumber \\
&& +~ \left[ \left( \left( \frac{\alpha^2}{4} - \frac{17\alpha}{24} 
- \frac{13}{8} \right) C_A^2 + C_A T_F\Nf \left( \frac{2}{3}\alpha + 1 \right) 
\right) \frac{1}{\epsilon^2} \right. \nonumber \\
&& \left. ~~~~~-~ \left( \left( \frac{\alpha^3}{4} + \frac{\alpha^2}{12} 
+ \frac{331\alpha}{144} - \frac{4115}{432} \right) C_A^2 + 2 C_F T_F \Nf 
- \frac{80}{27} T_F^2 \Nf^2 \right. \right. \nonumber \\
&& \left. \left. ~~~~~~~~~~~+~ \left( \frac{589}{54} - \frac{14\alpha}{9} 
+ \frac{\alpha^2}{3} \right) C_A T_F \Nf \right) \frac{1}{\epsilon} 
+ \left( \left( 16 \zeta(3)
- \frac{55}{3} \right) T_F\Nf C_F \right. \right. \nonumber \\
&& \left. \left. ~~~~~~~~~~~+~ \frac{400}{81} T_F^2 \Nf^2 
- \left( \frac{8659}{324} + \frac{20\alpha}{9} 
+ \frac{10\alpha^2}{9} + 8\zeta(3) \right) T_F \Nf C_A \right. \right. 
\nonumber \\
&& \left. \left. ~~~~~~~~~~~+~ \left( \frac{83105}{2592} 
+ \frac{701\alpha}{288} + \frac{365\alpha^2}{144} + \frac{11\alpha^3}{16} 
+ \frac{\alpha^4}{8} - 3\zeta(3) + 2 \alpha \zeta(3) \right) C_A^2 \right) 
\right] a^2 \nonumber \\ 
&& +~ \left[ \left( \left( \frac{403}{144} + \frac{47\alpha}{48} 
+ \frac{\alpha^2}{6} - \frac{\alpha^3}{8} \right) C_A^3 - \left( \frac{22}{9} 
+ \frac{5\alpha}{6} + \frac{\alpha^2}{3} \right) C_A^2 T_F \Nf  
+ \frac{4}{9} C_A T_F^2 \Nf^2 \right) \frac{1}{\epsilon^3} \right. \nonumber \\
&& \left. ~~~~+~ \left( \left( \frac{3\alpha^4}{16} + \frac{\alpha^3}{6} 
+ \frac{139\alpha^2}{96} - \frac{5287\alpha}{864} - \frac{2935}{216} \right) 
C_A^3 \right. \right. \nonumber \\
&& \left. \left. ~~~~~~~~~~~+~ \left( \frac{1643}{108} + \frac{953\alpha}{108} 
- \frac{7\alpha^2}{12} + \frac{\alpha^3}{3} \right) C_A^2 T_F \Nf 
- \left( \frac{110}{27} + \frac{80\alpha}{27} \right) C_A T_F^2 \Nf^2 
\right. \right. \nonumber \\
&& \left. \left. ~~~~~~~~~~~-~ \frac{8}{9} C_F T_F^2 \Nf^2 + \left( 
\frac{31}{9} + \alpha \right) C_A C_F T_F \Nf  \right) \frac{1}{\epsilon^2} 
\right. \nonumber \\
&& \left. ~~~~+~ \left( \left( \frac{88391}{972} - \frac{19595\alpha}{576} 
+ \frac{1021\alpha^2}{1728} - \frac{235\alpha^3}{96} - \frac{61\alpha^4}{96}
- \frac{5\alpha^5}{32} \right. \right. \right. \nonumber \\
&& \left. \left. \left. ~~~~~~~~~~~~~-~ \left( \frac{33\alpha^2}{16} 
- \frac{85\alpha}{12} + \frac{107}{16} \right) \zeta(3) \right) C_A^3 
\right. \right. \nonumber \\ 
&& \left. \left. ~~~~~~~~~-~ \left( \frac{85831}{648} - \frac{217\alpha}{8} 
+ \frac{485\alpha^2}{216} - \frac{3\alpha^3}{4} + \frac{\alpha^4}{6} 
+ \frac{22}{3} \zeta(3) - \frac{16\alpha}{3} \zeta(3) \right) C_A^2 T_F \Nf  
\right. \right. \nonumber \\ 
&& \left. \left. ~~~~~~~~~-~ \left( \frac{2441}{54} - \frac{52\alpha}{3} 
+ \frac{\alpha^2}{2} - \frac{80}{3} \zeta(3) + 16 \alpha \zeta(3) \right) 
C_A C_F T_F \Nf \right. \right. \nonumber \\ 
&& \left. \left. ~~~~~~~~~+~ \left( \frac{1477}{27} - \frac{40\alpha}{9} 
+ \frac{40\alpha^2}{27} + \frac{32}{3} \zeta(3) \right) C_A T_F^2 \Nf^2 
\right. \right. \nonumber \\
&& \left. \left. ~~~~~~~~~+~ \left( \frac{824}{27} - \frac{64}{3} \zeta(3) 
\right) C_F T_F^2 \Nf^2 + \frac{2}{3} C_F^2 T_F \Nf 
- \frac{1600}{243} T_F^3 \Nf^3 \right) \frac{1}{\epsilon} \right. \nonumber \\
&& \left. ~~~~+~ \left( \left( \left( 252 + 16 \alpha + 8 \alpha^2 \right) 
\zeta(3) - 12 \zeta(4) + 80 \zeta(5) - \frac{128819}{324} \right. \right.
\right. \nonumber \\
&& \left. \left. \left. ~~~~~~~~~-~ \frac{55\alpha}{3} 
- \frac{55\alpha^2}{6} \right) C_A C_F T_F \Nf + \left( \frac{286}{9} 
+ \frac{296}{3} \zeta(3) - 160 \zeta(5) \right) C_F^2 T_F \Nf \right. \right.
\nonumber \\
&& \left. \left. ~~~~~~~~~-~ \left( \frac{2080363}{3888} 
+ \frac{6115\alpha}{216} + \frac{10993\alpha^2}{432} + \frac{55\alpha^3}{12} 
+ \frac{5\alpha^4}{6} \right. \right. \right. \nonumber \\
&& \left. \left. \left. ~~~~~~~~~~~~~~~+~ \left( \frac{469}{6} 
+ \frac{430\alpha}{9} + \frac{23\alpha^2}{6} \right) \zeta(3) - 9 \zeta(4) 
- \frac{160}{3} \zeta(5) \right) C_A^2 T_F \Nf \right. \right. \nonumber \\
&& \left. \left. ~~~~~~~~~+~ \left( \frac{14002}{81} - \frac{416}{3} \zeta(3) 
\right) C_F T_F^2 \Nf^2 - \frac{8000}{729} T_F^3 \Nf^3 \right. \right.
\nonumber \\
&& \left. \left. ~~~~~~~~~+~ \left( \frac{12043}{81} + \frac{152\alpha}{27} 
+ \frac{100\alpha^2}{27} + 64 \zeta(3) + \frac{64\alpha}{9} \zeta(3) \right) 
C_A T_F^2 \Nf^2 \right. \right. \nonumber \\
&& \left. \left. ~~~~~~~~~+~ \left( \frac{44961125}{93312} 
+ \frac{14939\alpha}{432} 
+ \frac{125759\alpha^2}{3456} + \frac{497\alpha^3}{48} + \frac{233\alpha^4}{64}
+ \frac{45\alpha^5}{64} + \frac{5\alpha^6}{64} \right. \right. \right. 
\nonumber \\
&& \left. \left. \left. ~~~~~~~~~~~~~~~-~ \left( \frac{1937}{24} 
- \frac{15431\alpha}{288} - \frac{257\alpha^2}{96} - \frac{91\alpha^3}{96} 
+ \frac{13\alpha^4}{96} \right) \zeta(3) \right. \right. \right. \nonumber \\
&& \left. \left. \left. ~~~~~~~~~~~~~~~-~ \left( \frac{7025}{192} 
+ \frac{115\alpha}{8} + \frac{385\alpha^2}{96} - \frac{5\alpha^3}{24} 
- \frac{35\alpha^4}{192} \right) \zeta(5) \right. \right. \right. \nonumber \\
&& \left. \left. \left. ~~~~~~~~~~~~~~~-~ \left( \frac{9}{32} 
+ \frac{3\alpha}{8} + \frac{3\alpha^2}{32} \right) \zeta(4) \right) C_A^3 
\right) \right] a^3 ~+~ O(a^4) \\ 
Z^{\mbox{\footnotesize{RI$^\prime$}}}_\alpha &=& 1 ~+~ O(a^4)  
\end{eqnarray} 
\begin{eqnarray} 
Z^{\mbox{\footnotesize{RI$^\prime$}}}_c &=& 1 ~+~ \left( \left( \frac{3}{4} 
- \frac{\alpha}{4} \right) \frac{1}{\epsilon} + 1 \right) C_A a ~+~ 
\left[ \left( \left( \frac{3\alpha^2}{32} - \frac{35}{32}
\right) C_A^2 + \frac{1}{2} C_A T_F \Nf \right) \frac{1}{\epsilon^2} \right. 
\nonumber \\ 
&& \left. ~~~~-~ \left( \left( \frac{\alpha^3}{16} + \frac{\alpha^2}{8} 
+ \frac{257\alpha}{288} - \frac{167}{96} \right) C_A^2 + \left( \frac{5}{12} 
- \frac{5\alpha}{9} \right) C_A T_F \Nf \right) \frac{1}{\epsilon} 
\right. \nonumber \\
&& \left. ~~~~+~ \left( \left( \frac{1943}{192} - \frac{7\alpha}{64} 
+ \frac{3\alpha^2}{8} - \left( \frac{15}{16} - \frac{3\alpha}{8} 
+ \frac{3\alpha^2}{16} \right) \zeta(3) \right) C_A^2 
- \frac{95}{24} C_A T_F \Nf \right) \right] a^2 \nonumber \\ 
&& +~ \left[ \left( \left( \frac{2765}{1152} + \frac{35\alpha}{384} 
- \frac{9\alpha^2}{128} - \frac{5\alpha^3}{128} \right) C_A^3 
- C_A^2 T_F\Nf \left( \frac{149}{72} + \frac{\alpha}{24} \right) 
+ \frac{4}{9} C_A T_F^2 \Nf^2 \right) \frac{1}{\epsilon^3} \right. \nonumber \\
&& \left. ~~~~+~ \left( \left( \frac{3\alpha^4}{64} + \frac{11\alpha^3}{96} 
+ \frac{269\alpha^2}{384} + \frac{5\alpha}{96} - \frac{19367}{3456} \right) 
C_A^3 + C_A C_F T_F \Nf \right. \right. \nonumber \\
&& \left. \left. ~~~~~~~~~~~~+~ \left( \frac{1621}{432} - \frac{\alpha}{48} 
- \frac{5\alpha^2}{12} \right) C_A^2 T_F \Nf - \frac{10}{27} C_A T_F^2 \Nf^2 
\right) \frac{1}{\epsilon^2} \right. \nonumber \\
&& \left. ~~~~+~ \left( \left( \frac{241171}{20736} 
- \frac{117809\alpha}{10368} - \frac{1015\alpha^2}{2304} 
- \frac{919\alpha^3}{1152} - \frac{11\alpha^4}{64} - \frac{\alpha^5}{32}
\right. \right. \right. \nonumber \\
&& \left. \left. \left. ~~~~~~~~~~~+~ \left( \frac{3\alpha^3}{64} 
- \frac{45\alpha^2}{64} + \frac{89\alpha}{64} - \frac{39}{64} \right) \zeta(3) 
\right) C_A^3 \right. \right. \nonumber \\ 
&& \left. \left. ~~~~~~~~~~~-~ \left( \frac{9551}{2592} 
- \frac{21899\alpha}{2592} - \frac{5\alpha^2}{9} - \frac{5\alpha^3}{18}  
+ 3\zeta(3) - 2 \alpha \zeta(3) \right) C_A^2 T_F \Nf \right. \right. 
\nonumber \\ 
&& \left. \left. ~~~~~~~~~~~-~ \left( \frac{15}{4} - \frac{55\alpha}{12}
- 4 \zeta(3) + 4 \alpha \zeta(3) \right) C_A C_F T_F \Nf \right. \right.
\nonumber \\
&& \left. \left. ~~~~~~~~~~~-~ \left( \frac{35}{81} + \frac{100\alpha}{81}
\right) C_A T_F^2 \Nf^2 \right) \frac{1}{\epsilon} \right. \nonumber \\
&& \left. ~~~~+~ \left( \left( \frac{1082353}{7776} - \frac{313\alpha}{768}
+ \frac{253\alpha^2}{48} + \frac{989\alpha^3}{768} + \frac{3\alpha^4}{16}
\right. \right. \right. \nonumber \\
&& \left. \left. \left. ~~~~~~~~~~~-~ \left( \frac{13483}{576} 
- \frac{589\alpha}{64} + \frac{509\alpha^2}{192} + \frac{29\alpha^3}{64} 
+ \frac{3\alpha^4}{32} \right) \zeta(3) \right. \right. \right. \nonumber \\
&& \left. \left. \left. ~~~~~~~~~~~+~ \left( \frac{9}{64} + \frac{3\alpha}{16} 
+ \frac{3\alpha^2}{64} \right) \zeta(4) - \left( \frac{65}{32} 
+ \frac{65\alpha}{32} - \frac{35\alpha^2}{32} + \frac{5\alpha^3}{32} \right) 
\zeta(5) \right) C_A^3 \right. \right. \nonumber \\
&& \left. \left. ~~~~~~~~~~~+~ \left( 22 \zeta(3) + 6 \zeta(4) - \frac{899}{24}
\right) C_A C_F T_F \Nf 
+ \left( \frac{5161}{486} + \frac{8}{9} \zeta(3) \right) C_A T_F^2 \Nf^2
\right. \right. \nonumber \\
&& \left. \left. ~~~~~~~~~~~-~ \left( \frac{165637}{1944} - \frac{13\alpha}{8} 
+ \frac{5\alpha^2}{3} + \left( \frac{29}{9} + 3 \alpha - \frac{5\alpha^2}{6} 
\right) \zeta(3) \right. \right. \right. \nonumber \\
&& \left. \left. \left. ~~~~~~~~~~~~~~~~~~+~ \frac{9}{2} \zeta(4) \right) 
C_A^2 T_F \Nf \right) \right] a^3 ~+~ O(a^4) 
\end{eqnarray} 
\begin{eqnarray} 
Z^{\mbox{\footnotesize{RI$^\prime$}}}_\psi &=& 1 ~-~ 
\left ( \frac{\alpha C_F}{\epsilon} + \alpha C_F \right) a ~+~ \left[ \left( 
C_F C_A \left( \frac{\alpha^2}{4} + \frac{3\alpha}{4} \right) 
+ \frac{\alpha^2}{2} C_F^2 \right) \frac{1}{\epsilon^2} \right. \nonumber \\ 
&& \left. ~~-~ \left( \left( \frac{\alpha^4}{4} + \frac{5\alpha^3}{8} 
+ \frac{\alpha^2}{8} + \frac{133\alpha}{36} + \frac{25}{8} \right) C_F C_A 
\right. \right. \nonumber \\
&& \left. \left. ~~~~~~~~-~ \left( 1 + \frac{20\alpha}{9} \right) 
C_F T_F \Nf ~-~ \left( \frac{3}{4} + \alpha^2 \right) C_F^2 \right) 
\frac{1}{\epsilon} \right. \nonumber \\
&& \left. ~~~~~~~~+~ \left( \left( 3\zeta(3) + 3 \alpha \zeta(3) - \frac{41}{4} 
- \frac{331\alpha}{36} - \frac{13\alpha^2}{8} - \frac{\alpha^4}{4} \right) 
C_F C_A \right. \right. \nonumber \\
&& \left. \left. ~~~~~~~~~~~~~~+~ \left( \frac{7}{2} + \frac{20\alpha}{9} 
\right) C_F T_F \Nf + \left( \frac{5}{8} + \alpha^2 \right) C_F^2 \right) 
\right] a^2 \nonumber \\ 
&& +\, \left[ \left( \frac{\alpha}{3} C_A C_F T_F \Nf - \left( 
\frac{3\alpha^2}{4} + \frac{\alpha^3}{4} \right) C_A C_F^2 - \left( 
\frac{31\alpha}{24} + \frac{3\alpha^2}{8} + \frac{\alpha^3}{12} \right) 
C_A^2 C_F - \frac{\alpha}{6} C_F^3 \right) \frac{1}{\epsilon^3} \right. 
\nonumber \\ 
&& \left. ~~~~+\, \left( \frac{8}{9} C_F T_F^2 \Nf^2 - \left( \frac{3\alpha}{4} 
+ \frac{\alpha^3}{2} \right) C_F^3 + \left( \frac{2}{3} - \alpha 
- \frac{20\alpha^2}{9} \right) C_F^2 T_F \Nf \right. \right. \nonumber \\
&& \left. \left. ~~~~~~~~~-~ \left( \frac{47}{9} + \frac{8\alpha}{3} 
+ \frac{10\alpha^2}{9} \right) C_A C_F T_F \Nf \right. \right. \nonumber \\
&& \left. \left. ~~~~~~~~~+~ \left( \frac{25\alpha}{8} + \frac{53\alpha^2}{18} 
+ \frac{3\alpha^3}{8} + \frac{\alpha^4}{4} - \frac{11}{6} \right) C_A C_F^2 
\right. \right. \nonumber \\
&& \left. \left. ~~~~~~~~~+~ \left( \frac{275}{36} + \frac{81\alpha}{16} 
+ \frac{89\alpha^2}{36} + \frac{9\alpha^3}{16} + \frac{\alpha^4}{8} \right) 
C_A^2 C_F \right) \frac{1}{\epsilon^2} \right.  \nonumber \\  
&& \left. ~~~~+~ \left( \frac{287}{27} + \frac{4919\alpha}{162}
+ \frac{25\alpha^2}{9} + \frac{10\alpha^3}{9} + 8 \alpha \zeta(3) \right) C_A
C_F T_F \Nf \right. \nonumber \\
&& \left. \left. ~~~~~~~~~~-~ \left( \frac{20}{27} + \frac{400\alpha}{81} 
\right) C_F T_F^2 \Nf^2 - \left( \frac{1}{2} + \frac{11\alpha}{8} + \alpha^3 
\right) C_F^3 \right. \right. \nonumber \\
&& \left. \left. ~~~~~~~~~~-~ \left( 1 - \frac{83\alpha}{6} 
+ \frac{40\alpha^2}{9} + 16 \alpha \zeta(3) \right) C_F^2 T_F \Nf 
\right. \right. \nonumber \\ 
&& \left. \left. ~~~~~~~~~~-~ \left( \frac{\alpha^5}{8} + \frac{3\alpha^4}{4} 
+ \frac{217\alpha^3}{72} + \frac{289\alpha^2}{72} + \frac{48595\alpha}{1296} 
+ \frac{9155}{432} \right. \right. \right. \nonumber \\
&& \left. \left. \left. ~~~~~~~~~~~~~~~~-~ \left( \frac{23}{8} 
+ \frac{11\alpha}{4} - \frac{17\alpha^2}{8} \right) \zeta(3) \right) C_A^2 C_F 
\right. \right. \nonumber \\
&& \left. ~~~~~~~~~~~+~ \left( \frac{143}{12} + \frac{107\alpha}{8} 
+ \frac{116\alpha^2}{9} + \frac{9\alpha^3}{4} + \frac{\alpha^4}{2} 
- \left( 4 + 3 \alpha + 3 \alpha^2 \right) \zeta(3) \right) C_A C_F^2 \right) 
\frac{1}{\epsilon} \nonumber \\  
&& \left. ~~~~+~ \left( \left( \frac{11887}{81} + \frac{42185\alpha}{648}
+ \frac{65\alpha^2}{9} + \frac{10\alpha^3}{9} - \frac{52}{3} \zeta(3) 
- \frac{20\alpha}{3} \zeta(3) \right) C_F C_A T_F \Nf \right.
\right. \nonumber \\
&& \left. \left. ~~~~~~~~~~+~ \left( \frac{79}{6} + \frac{77\alpha}{6} 
- \frac{40\alpha^2}{9} - 16\zeta(3) - 16 \alpha \zeta(3) \right) 
C_F^2 T_F \Nf \right. \right. \nonumber \\
&& \left. \left. ~~~~~~~~~~-~ \left( \frac{1570}{81} + \frac{400\alpha}{81} 
\right) C_F T_F^2 \Nf^2 \right. \right. \nonumber \\
&& \left. ~~~~~~~~~~+~ \left( \left( \frac{3139}{24} + \frac{553\alpha}{12}  
+ \frac{35\alpha^2}{8} + \frac{13\alpha^3}{12} \right) \zeta(3) 
+ \left( \frac{69}{16} - \frac{3\alpha}{8} - \frac{3\alpha^2}{16} \right) 
\zeta(4) \right. \right. \nonumber \\ 
&& \left. \left. ~~~~~~~~~~~~~~~~~-~ \left( \frac{165}{4} + \frac{5\alpha}{2} 
+ \frac{5\alpha^2}{4} \right) \zeta(5) - \frac{159257}{648} 
- \frac{615193\alpha}{5184} \right. \right. \nonumber \\
&& \left. \left. ~~~~~~~~~~~~~~~~~-~ \frac{13849\alpha^2}{576} 
- \frac{1091\alpha^3}{144} - \frac{5\alpha^4}{4} - \frac{\alpha^5}{8} \right) 
C_F C_A^2 \right. \nonumber \\
&& \left. \left. ~~~~~~~~~~+~ \left( \frac{997}{24} + \frac{33\alpha}{2} 
+ \frac{152\alpha^2}{9} + \frac{27\alpha^3}{8} + \frac{\alpha^4}{2} 
- \left( 44 - 11 \alpha + 6 \alpha^2 + \alpha^3 \right) \zeta(3) 
\right. \right. \right. \nonumber \\
&& \left. \left. \left. ~~~~~~~~~~~~~~~~-~ 6 \zeta(4) + 20 ( 1 - \alpha ) 
\zeta(5) \frac{}{} \! \right) C_F^2 C_A \right. \right. \nonumber \\
&& \left. \left. ~~~~~~~~~~+~ \left( \frac{73}{12} - \frac{17\alpha}{8} 
- \alpha^3 + \frac{2\alpha^3}{3} \zeta(3) \right) C_F^3 \right) \right] 
a^3 ~+~ O(a^4) 
\end{eqnarray} 
where $\mbox{Tr} \left( T^a T^b \right)$ $=$ $T_F \delta^{ab}$, $T^a T^a$ $=$ 
$C_F I$, $f^{acd} f^{bcd}$ $=$ $C_A \delta^{ab}$, $\zeta(n)$ is the Riemann 
zeta function, $a$ $=$ $g^2/(16\pi^2)$ and $g$ is the coupling constant 
appearing in the covariant derivative. At this point it is worth commenting on
the status of the variables $a$ and $\alpha$. As we have computed the 
renormalization constants for the RI$^\prime$ scheme they correspond to the 
RI$^\prime$ scheme coupling constant and covariant gauge parameter. When it is
necessary we will include a label on the variables to distinguish which scheme
they are defined in. As with other schemes they can be related to the 
corresponding $\MSbar$ variables which we will discuss later. Throughout this 
article when we quote any renormalization constant the scheme will be denoted 
on the renormalization constant itself and it will be understood that the 
variables will be in that scheme as well. For completeness we note,  
\begin{eqnarray} 
Z^{\mbox{\footnotesize{$\MSbar$}}}_\psi &=& 1 ~-~ \alpha C_F 
\frac{a}{\epsilon} ~+~ \left[ \left( C_F C_A  
\left( \frac{\alpha^2}{4} + \frac{3\alpha}{4} \right) + \frac{\alpha^2}{2} 
C_F^2 \right) \frac{1}{\epsilon^2} \right. \nonumber \\ 
&& \left. ~~~~~~~~~~~~~~~~~~~~~~~-~ \left( C_F C_A \left( \frac{\alpha^2}{8} 
+ \alpha + \frac{25}{8} \right) ~-~ C_F T_F \Nf ~-~ \frac{3}{4} C_F^2 \right) 
\frac{1}{\epsilon} \right] a^2 \nonumber \\ 
&& +~ \left[ \left( \frac{\alpha}{3} C_A C_F T_F \Nf - \left( 
\frac{3\alpha^2}{4} + \frac{\alpha^3}{4} \right) C_A C_F^2 \right. \right.
\nonumber \\
&& \left. \left. ~~~~-~ \left( \frac{31\alpha}{24} + \frac{3\alpha^2}{8} 
+ \frac{\alpha^3}{12} \right) C_A^2 C_F - \frac{\alpha^3}{6} C_F^3 \right) 
\frac{1}{\epsilon^3} \right. \nonumber \\ 
&& \left. ~~~~+~ \left( \frac{8}{9} C_F T_F^2 \Nf^2 - \frac{3\alpha}{4} C_F^3 
+ \left( \frac{2}{3} - \alpha \right) C_F^2 T_F \Nf 
- \left( \frac{47}{9} + \alpha \right) C_A C_F T_F \Nf \right. \right. 
\nonumber \\
&& \left. \left. ~~~~~~~~~~+~ \left( \frac{\alpha^3}{8} + \alpha^2 
+ \frac{25\alpha}{8} - \frac{11}{6} \right) C_A C_F^2 + \left( \frac{275}{36} 
+ \frac{73\alpha}{24} + \frac{3\alpha^2}{4} + \frac{\alpha^3}{8} \right) C_A^2 
C_F \right) \frac{1}{\epsilon^2} \right.  \nonumber \\  
&& \left. ~~~~+~ \left( -~ \frac{20}{27} C_F T_F^2 \Nf^2 - \frac{1}{2} C_F^3 
- C_F^2 T_F \Nf + \left( \frac{287}{27} + \frac{17\alpha}{12} \right) C_A C_F 
T_F \Nf \right. \right. \nonumber \\
&& \left. \left. ~~~~~~~~~~-~ \left( \frac{5\alpha^3}{48} 
+ \frac{13\alpha^2}{32} + \frac{263\alpha}{96} + \frac{9155}{432} 
- \left( \frac{23}{8} - \frac{\alpha}{4} - \frac{\alpha^2}{8} \right) \zeta(3) 
\right) C_A^2 C_F \right. \right. \nonumber \\
&& \left. \left. ~~~~~~~~~~+~ \left( \frac{143}{12} - 4 \zeta(3) \right) 
C_A C_F^2 \right) \frac{1}{\epsilon} \right] a^3 ~+~ O(a^4) 
\end{eqnarray} 
where the variables $a$ and $\alpha$ correspond to those of the $\MSbar$ 
scheme. There are several checks on these results. First, we have verified that
the correct three loop $\MSbar$ renormalization constants, 
\cite{25,26,27,28,29}, emerge with the programmes we have written prior to 
extracting the results for the RI$^\prime$ scheme. Second, the gauge parameter,
$\alpha$, does not get renormalized as in the $\MSbar$ scheme so that gauge 
invariance is not destroyed. Next, our three loop result for 
$Z^{\mbox{\footnotesize{RI$^\prime$}}}_\psi$ agrees with the four loop result 
of \cite{12} in the {\em Landau} gauge. They do not agree for non-zero
$\alpha$. The reason for this is that in the RI$^\prime$ scheme used in 
\cite{12} the values taken for $Z_A$ and $Z_\alpha$ were different from those 
given above. More specifically the $\MSbar$ expressions were used which only 
differ in the case $\alpha$ $\neq$ $0$. However, using the scheme adopted in
\cite{12} we have reproduced the expressions given there for all $\alpha$ which
again confirms the correctness of our programming. However, as we believe the 
full RI$^\prime$ scheme we have introduced here is more {\em natural} we will 
always present subsequent results with reference to the above renormalization 
constants. Indeed it is not inconceivable that at some point one would require 
the renormalization of an operator with gluon content in RI$^\prime$ and 
therefore one would require our $Z^{\mbox{\footnotesize{RI$^\prime$}}}_A$. 
Nevertheless all results should be consistent with \cite{12} in the Landau 
gauge, $\alpha$~$=$~$0$.  

Whilst the results for the field renormalizations are deduced from the 
two-point functions we have yet to establish their consistency. This can be 
verified by renormalizing several three-point functions to obtain the same 
coupling constant renormalizations. We have examined both the quark gluon and 
ghost gluon vertices. However, with the wave function renormalization 
containing a finite part at one loop the method of renormalizing the vertices 
in the RI$^\prime$ scheme cannot be the same as for the two-point functions. In
other words one cannot absorb a finite part from the three-point functions into
the definition of the coupling constant renormalization. For instance, at one 
loop the quark gluon vertex would give an $\Nf$ dependent finite part to that 
coupling constant renormalization constant. However, this $\Nf$ dependence 
cannot be matched in the gluon ghost vertex which is $\Nf$ independent at one 
loop. Although it is possible to follow this route and accommodate the problem 
of the finite parts in the three-point vertices it will lead to an MOM or 
$\MOMbar$ renormalization scheme which we are not considering here, \cite{7,8}.
Therefore, to define a three loop renormalization constant for $Z_g$ in the 
RI$^\prime$ scheme which is consistent with the Slavnov-Taylor identities we 
define the $Z_g$ renormalization for each vertex in an $\MSbar$ way. In other 
words we define $Z_g$ by only absorbing the infinities without removing any 
finite parts, aside from powers of $\ln(4\pi e^{-\gamma})$. With this we can 
consistently deduce the same coupling constant renormalization from both 
three-point functions. More concretely we have  
\begin{equation} 
\left. \lim_{\epsilon \rightarrow 0} \left[ 
Z^{\mbox{\footnotesize{RI$^\prime$}}}_\psi 
\left( Z^{\mbox{\footnotesize{RI$^\prime$}}}_A \right)^{\half}  
Z^{\mbox{\footnotesize{RI$^\prime$}}}_g G^{\mu \, a}_{A\bar{\psi}\psi}(p) 
\right] \right|_{p^2 \, = \, \mu^2} ~=~ 
G^{\mu \, a \, \, \mbox{\footnotesize{finite}}}_{A\bar{\psi}\psi} 
\end{equation} 
and 
\begin{equation} 
\left. \lim_{\epsilon \rightarrow 0} \left[ 
Z^{\mbox{\footnotesize{RI$^\prime$}}}_c 
\left( Z^{\mbox{\footnotesize{RI$^\prime$}}}_A \right)^{\half}  
Z^{\mbox{\footnotesize{RI$^\prime$}}}_g G_{A\bar{c}c}(p) \right] 
\right|_{p^2 \, = \, \mu^2} ~=~ 
G^{\mbox{\footnotesize{finite}}}_{A\bar{c}c} 
\end{equation} 
where $G^{\mbox{\footnotesize{finite}}}_i$ are treated as finite with respect
to $\epsilon$ and are {\em not} unity and we have omitted the overall structure
function from the ghost gluon Green's function. Though the tree part of each 
$G^{\mbox{\footnotesize{finite}}}_i$ is unity since we have first divided 
$G_i(p)$ by the bare coupling constant before rescaling as in the method of 
\cite{24}. To evaluate $Z^{\mbox{\footnotesize{RI$^\prime$}}}_g$ explicitly we 
have again used {\sc Mincer}, \cite{20}. However, to do this correctly the 
external momentum of one leg must be nullified as {\sc Mincer} can only be 
applied to two-point functions. To avoid potential spurious infrared infinities
arising in this case we have chosen to nullify an external quark or ghost leg 
respectively leaving $p$ as the overall external momentum. Consequently we find
from both three-point functions that  
\begin{eqnarray} 
Z^{\mbox{\footnotesize{RI$^\prime$}}}_g &=& 1 ~+~ 
\left( \frac{2}{3} T_F \Nf - \frac{11}{6} C_A \right) 
\frac{a}{\epsilon} ~+~ \left[ \left( \frac{121}{24}C_A^2 + \frac{2}{3} T_F^2
\Nf^2 - \frac{11}{3} C_A T_F \Nf \right) \frac{1}{\epsilon^2} \right. 
\nonumber \\ 
&& \left. +~ \left( C_F T_F \Nf + \frac{5}{3} C_A T_F \Nf - \frac{17}{6} 
C_A^2 \right) \frac{1}{\epsilon} \right] a^2 \nonumber \\  
&& +~ \left[ \left( \frac{605}{36} C_A^2 T_F \Nf - \frac{55}{9} C_A T_F^2 \Nf^2
+ \frac{20}{27} T_F^3 \Nf^3 - \frac{6655}{432} C_A^3 \right) 
\frac{1}{\epsilon^3} \right. \nonumber \\ 
&& \left. ~~~~+\, \left( \frac{22}{9} C_F T_F^2 \Nf^2 - \frac{121}{18} 
C_A C_F T_F \Nf - \frac{979}{54} C_A^2 T_F \Nf + \frac{110}{27} C_A T_F^2 \Nf^2
+ \frac{2057}{108} C_A^3 \! \right) \! \frac{1}{\epsilon^2} \right. 
\nonumber \\
&& \left. ~~~~+\, \left( \frac{205}{54} C_A C_F T_F \Nf 
- \frac{22}{27} C_F T_F^2 \Nf^2 + \frac{1415}{162} C_A^2 T_F \Nf 
\right. \right. \nonumber \\ 
&& \left. \left. ~~~~~~~~~~-~ \frac{79}{81} C_A T_F^2 \Nf^2 
- \frac{1}{3} C_F^2 T_F \Nf - \frac{2857}{324} C_A^3 \right) \frac{1}{\epsilon}
\right] a^3 ~+~ O(a^4) 
\label{Zgris} 
\end{eqnarray}  
which is the same as in the $\MSbar$ scheme, \cite{29,28}, though it could only
have differed in the three loop term which is scheme dependent. Thus to this 
order the RI$^\prime$ and $\MSbar$ $\beta$-functions coincide. In our 
conventions we have, for all $\alpha$,  
\begin{eqnarray} 
\beta^{\mbox{\footnotesize{RI$^\prime$}}} 
(a_{\mbox{\footnotesize{RI$^\prime$}}}) &=& -~ \left[ \frac{11}{3} C_A 
- \frac{4}{3} T_F \Nf \right] a^2_{\mbox{\footnotesize{RI$^\prime$}}} ~-~ 
\left[ \frac{34}{3} C_A^2 - 4 C_F T_F \Nf - \frac{20}{3} C_A T_F \Nf \right]
a^3_{\mbox{\footnotesize{RI$^\prime$}}} \nonumber \\  
&& +~ \left[ 2830 C_A^2 T_F \Nf - 2857 C_A^3 + 1230 C_A C_F T_F \Nf 
- 316 C_A T_F^2 \Nf^2 \right. \nonumber \\ 
&& \left. ~~~~~-~ 108 C_F^2 T_F \Nf - 264 C_F T_F^2 \Nf^2 \right] 
\frac{a^4_{\mbox{\footnotesize{RI$^\prime$}}}}{54} ~+~ 
O( a^5_{\mbox{\footnotesize{RI$^\prime$}}} ) 
\label{betaris} 
\end{eqnarray} 
where we have explicitly indicated the scheme of the coupling constant as a 
subscript. Given that the three loop term of the $\beta$-function can be 
different we have constructed the relation between the coupling constants of 
the RI$^\prime$ and $\MSbar$ schemes explicitly. To do this we parallel the 
approach of \cite{30} where the same procedure was followed to establish the 
relation between various $\MOMbar$ coupling constants and the $\MSbar$ 
coupling. Since we have renormalized two different three-point functions we 
have to check that both give equivalent results and are consistent with the 
Slavnov-Taylor identities. We define 
\begin{equation} 
a_{\mbox{\footnotesize{RI$^\prime$}}}(\mu) ~=~  
a_{\mbox{\footnotesize{$\MSbar$}}}(\mu) \left. \left[ 
\frac{1}{\Pi_T^{\mbox{\footnotesize{$\MSbar$~finite}}}(p) 
\left(\Sigma^{\mbox{\footnotesize{$\MSbar$~finite}}}_c(p) \right)^2}  
\left[ \frac{G^{\mbox{\footnotesize{$\MSbar$~finite}}}_{A\bar{c}c}}  
{G^{\mbox{\footnotesize{RI$^\prime$~finite}}}_{A\bar{c}c}} \right]^2 \right] 
\right|_{p^2 \, = \, \mu^2}
\label{qqvdef}
\end{equation}  
and 
\begin{equation} 
a_{\mbox{\footnotesize{RI$^\prime$}}}(\mu) ~=~ 
a_{\mbox{\footnotesize{$\MSbar$}}}(\mu) \left. \left[ 
\frac{1}{\Pi_T^{\mbox{\footnotesize{$\MSbar$~finite}}}(p)
\left(\Sigma^{\mbox{\footnotesize{$\MSbar$~finite}}}_\psi(p) \right)^2} \left[ 
\frac{G^{(1) \, \mbox{\footnotesize{$\MSbar$~finite}}}_{A\bar{\psi}\psi}} 
{G^{(1) \, \mbox{\footnotesize{RI$^\prime$~finite}}}_{A\bar{\psi}\psi}} 
\right]^2 \right] \right|_{p^2 \, = \, \mu^2}
\label{accdef}
\end{equation}  
where the Green's functions on the right hand side are the finite expressions. 
Moreover, since we are concerned with the finite part of the Green's function 
after renormalization  $G^{(1)}_{A\bar{\psi}\psi}$ corresponds to a particular 
Lorentz projection of the full Green's function 
$G^{\mu \, a}_{A\bar{\psi}\psi}$, \cite{30}. In particular, for the momentum 
routing we are considering, we have defined, \cite{30},  
\begin{equation} 
G^{\mu \, a}_{A\bar{\psi}\psi}(p) ~=~ T^a \left[ G^{(1)}_{A\bar{\psi}\psi}(p) 
\gamma^\mu ~+~   G^{(2)}_{A\bar{\psi}\psi}(p) \left( \gamma^\mu 
- \frac{\pslash p^\mu}{p^2} \right) \right] 
\end{equation}  
where 
\begin{eqnarray} 
T^a G^{(1)}_{A\bar{\psi}\psi}(p) &=& \frac{1}{4p^2} \mbox{tr} \left[ p_\mu 
\pslash G^{\mu \, a}_{A\bar{\psi}\psi}(p) \right] \nonumber \\ 
T^a G^{(2)}_{A\bar{\psi}\psi}(p) &=& \frac{1}{4(d-1)} \left[ \mbox{tr} \left(  
\gamma_\mu G^{\mu \, a}_{A\bar{\psi}\psi}(p) \right) ~-~ d \, \mbox{tr} \left( 
\frac{\pslash p_\mu}{p^2} G^{\mu \, a}_{A\bar{\psi}\psi}(p) \right) \right] ~. 
\end{eqnarray} 
For the ghost gluon vertex we do not need to take into account several 
projections since there is only one Lorentz structure for that vertex with one
external momentum nullified. In addition to computing  
$a_{\mbox{\footnotesize{RI$^\prime$}}}$ as a function of  
$a_{\mbox{\footnotesize{$\MSbar$}}}$ for both vertices we will also need the 
relation of the covariant gauge parameter in one scheme with that in the other
since, for instance, 
$G^{\mbox{\footnotesize{RI$^\prime$~finite}}}_{A\bar{c}c}$ depends on  
$a_{\mbox{\footnotesize{RI$^\prime$}}}$ and 
$\alpha_{\mbox{\footnotesize{RI$^\prime$}}}$. This is achieved by, \cite{30},
\begin{equation}
\alpha_{\mbox{\footnotesize{RI$^\prime$}}} ~=~ 
\frac{Z^{\mbox{\footnotesize{RI$^\prime$}}}_A}
{Z^{\mbox{\footnotesize{$\MSbar$}}}_A} 
\alpha_{\mbox{\footnotesize{$\MSbar$}}} ~.  
\label{alphadef}
\end{equation} 
By solving (\ref{qqvdef}), (\ref{accdef}) and (\ref{alphadef}) iteratively we 
have determined the relationships between the coupling constants and covariant 
gauge parameters in both schemes at three loops. From both the vertices we find
\begin{equation}   
a_{\mbox{\footnotesize{RI$^\prime$}}} ~=~ 
a_{\mbox{\footnotesize{$\MSbar$}}} ~+~ O \left( 
a_{\mbox{\footnotesize{$\MSbar$}}}^5 \right) 
\end{equation}  
for all gauges and for the covariant gauge parameter  
\begin{eqnarray}
\alpha_{\mbox{\footnotesize{RI$^\prime$}}} 
&=& \left[ 1 + \left( \left( - 9 \alpha_{\mbox{\footnotesize{$\MSbar$}}}^2 
- 18 \alpha_{\mbox{\footnotesize{$\MSbar$}}} - 97 \right) C_A + 80 T_F \Nf 
\right) \frac{a_{\mbox{\footnotesize{$\MSbar$}}}}{36} \right. \nonumber \\ 
&& \left. ~+~ \left( \left( 18 \alpha_{\mbox{\footnotesize{$\MSbar$}}}^4 
- 18 \alpha_{\mbox{\footnotesize{$\MSbar$}}}^3 
+ 190 \alpha_{\mbox{\footnotesize{$\MSbar$}}}^2 
- 576 \zeta(3) \alpha_{\mbox{\footnotesize{$\MSbar$}}} 
+ 463 \alpha_{\mbox{\footnotesize{$\MSbar$}}} + 864 \zeta(3) - 7143 \right) 
C_A^2 \right. \right. \nonumber \\ 
&& \left. \left. ~~~~~~~+~ \left( -~ 320 
\alpha_{\mbox{\footnotesize{$\MSbar$}}}^2 
- 320 \alpha_{\mbox{\footnotesize{$\MSbar$}}} + 2304 \zeta(3) 
+ 4248 \right) C_A T_F \Nf \right. \right. \nonumber \\ 
&& \left. \left. ~~~~~~~+~ \left( -~ 4608 \zeta(3) + 5280 \right) C_F T_F \Nf 
\right) \frac{a^2_{\mbox{\footnotesize{$\MSbar$}}}}{288} \right. \nonumber \\
&& \left. ~+~ \left( \left( ~-~ 486 \alpha_{\mbox{\footnotesize{$\MSbar$}}}^6 
+ 1944 \alpha_{\mbox{\footnotesize{$\MSbar$}}}^5 
+ 4212 \zeta(3) \alpha_{\mbox{\footnotesize{$\MSbar$}}}^4 
- 5670 \zeta(5) \alpha_{\mbox{\footnotesize{$\MSbar$}}}^4 
- 18792 \alpha_{\mbox{\footnotesize{$\MSbar$}}}^4 \right. \right. \right.   
\nonumber \\
&& \left. \left. \left. ~~~~~~~~+~ 48276 \zeta(3) 
\alpha_{\mbox{\footnotesize{$\MSbar$}}}^3 
- 6480 \zeta(5) \alpha_{\mbox{\footnotesize{$\MSbar$}}}^3 
- 75951 \alpha_{\mbox{\footnotesize{$\MSbar$}}}^3 
- 52164 \zeta(3) \alpha_{\mbox{\footnotesize{$\MSbar$}}}^2 
\right. \right. \right. \nonumber \\
&& \left. \left. \left. ~~~~~~~~+~ 2916 \zeta(4) 
\alpha_{\mbox{\footnotesize{$\MSbar$}}}^2 
+ 124740 \zeta(5) \alpha_{\mbox{\footnotesize{$\MSbar$}}}^2 
+ 92505 \alpha_{\mbox{\footnotesize{$\MSbar$}}}^2
- 1303668 \zeta(3) \alpha_{\mbox{\footnotesize{$\MSbar$}}} 
\right. \right. \right. \nonumber \\
&& \left. \left. \left. ~~~~~~~~+~ 11664 \zeta(4) 
\alpha_{\mbox{\footnotesize{$\MSbar$}}} 
+ 447120 \zeta(5) \alpha_{\mbox{\footnotesize{$\MSbar$}}} 
+ 354807 \alpha_{\mbox{\footnotesize{$\MSbar$}}} 
+ 2007504 \zeta(3) 
\right. \right. \right. \nonumber \\
&& \left. \left. \left. ~~~~~~~~+~ 8748 \zeta(4) + 1138050 \zeta(5) 
- 10221367 \right) C_A^3 \right. \right. \nonumber \\
&& \left. \left. ~~~~~~~+~ \left( 
12960 \alpha_{\mbox{\footnotesize{$\MSbar$}}}^4 
- 8640 \alpha_{\mbox{\footnotesize{$\MSbar$}}}^3 
- 129600 \zeta(3) \alpha_{\mbox{\footnotesize{$\MSbar$}}}^2 
- 147288 \alpha_{\mbox{\footnotesize{$\MSbar$}}}^2
+ 698112 \zeta(3) \alpha_{\mbox{\footnotesize{$\MSbar$}}} 
\right. \right. \right. \nonumber \\
&& \left. \left. \left. ~~~~~~~~~~~~-~ 312336 
\alpha_{\mbox{\footnotesize{$\MSbar$}}} + 1505088 \zeta(3) - 279936 \zeta(4)
\right. \right. \right. \nonumber \\
&& \left. \left. \left. ~~~~~~~~~~~~-~ 1658880 \zeta(5) + 9236488 \right) 
C_A^2 T_F \Nf  \right. \right. \nonumber \\
&& \left. \left. ~~~~~~~+~ \left( 248832 \zeta(3) 
\alpha_{\mbox{\footnotesize{$\MSbar$}}}^2 
- 285120 \alpha_{\mbox{\footnotesize{$\MSbar$}}}^2 
+ 248832 \zeta(3) \alpha_{\mbox{\footnotesize{$\MSbar$}}} 
- 285120 \alpha_{\mbox{\footnotesize{$\MSbar$}}} \right. \right. \right.  
\nonumber \\
&& \left. \left. \left. ~~~~~~~~~~~~-~ 5156352 \zeta(3) + 373248 \zeta(4)
- 2488320 \zeta(5) + 9293664 \right) C_A C_F T_F \Nf \right. \right. 
\nonumber \\
&& \left. \left. ~~~~~~~+~ \left( ~-~ 38400 
\alpha_{\mbox{\footnotesize{$\MSbar$}}}^2 
- 221184 \zeta(3) \alpha_{\mbox{\footnotesize{$\MSbar$}}} 
+ 55296 \alpha_{\mbox{\footnotesize{$\MSbar$}}}
\right. \right. \right. \nonumber \\
&& \left. \left. \left. ~~~~~~~~~~~~~~-~ 884736 \zeta(3) - 1343872 \right) 
C_A T_F^2 \Nf^2 \right. \right. \nonumber \\
&& \left. \left. ~~~~~~~+~ \left( ~-~ 3068928 \zeta(3) + 4976640 \zeta(5) 
- 988416 \right) C_F^2 T_F \Nf \right. \right. \nonumber \\
&& \left. \left. ~~~~~~~+~ \left( 2101248 \zeta(3) - 2842368 \right) 
C_F T_F^2 \Nf^2 \right) 
\frac{a^3_{\mbox{\footnotesize{$\MSbar$}}}}{31104} \right] 
\alpha_{\mbox{\footnotesize{$\MSbar$}}} ~+~ O \left( 
a^4_{\mbox{\footnotesize{$\MSbar$}}} \right) ~.  
\end{eqnarray}
The former expression is consistent with the three loop RI$^\prime$ 
$\beta$-function being equivalent to the $\MSbar$ one for arbitrary covariant 
gauge. Moreover, since the three loop term of the transformation is also absent
this implies that the {\em four} loop RI$^\prime$ $\beta$-function is also 
equivalent to the {\em four} loop $\MSbar$ $\beta$-function in all gauges. The
non-trivial relation between the gauge parameters will be crucial in carrying
out checks on the renormalization group functions.   

We have calculated the renormalization group functions for the various wave 
function renormalizations directly from the renormalization constants 
themselves. In particular we used  
\begin{eqnarray} 
\gamma^{\mbox{\footnotesize{RI$^\prime$}}}_A(a) &=& \beta(a) \frac{\partial 
\ln Z^{\mbox{\footnotesize{RI$^\prime$}}}_A}{\partial a} ~+~ \alpha 
\gamma^{\mbox{\footnotesize{RI$^\prime$}}}_\alpha(a) \frac{\partial \ln 
Z^{\mbox{\footnotesize{RI$^\prime$}}}_A}{\partial \alpha} \nonumber \\
\gamma^{\mbox{\footnotesize{RI$^\prime$}}}_\alpha(a) &=& \left[ \beta(a) 
\frac{\partial \ln Z^{\mbox{\footnotesize{RI$^\prime$}}}_\alpha} 
{\partial a} ~-~ \gamma^{\mbox{\footnotesize{RI$^\prime$}}}_A(a) \right] 
\left[ 1 ~-~ \alpha \frac{\partial \ln 
Z^{\mbox{\footnotesize{RI$^\prime$}}}_\alpha}{\partial \alpha} \right]^{-1} 
\nonumber \\
\gamma^{\mbox{\footnotesize{RI$^\prime$}}}_\psi(a) &=& \beta(a) \frac{\partial 
\ln Z^{\mbox{\footnotesize{RI$^\prime$}}}_\psi}{\partial a} ~+~ \alpha 
\gamma^{\mbox{\footnotesize{RI$^\prime$}}}_\alpha(a) \frac{\partial \ln 
Z^{\mbox{\footnotesize{RI$^\prime$}}}_\psi}{\partial \alpha} \nonumber \\
\gamma^{\mbox{\footnotesize{RI$^\prime$}}}_c(a) &=& \beta(a) \frac{\partial 
\ln Z^{\mbox{\footnotesize{RI$^\prime$}}}_c}{\partial a} ~+~ \alpha 
\gamma^{\mbox{\footnotesize{RI$^\prime$}}}_\alpha(a) \frac{\partial \ln 
Z^{\mbox{\footnotesize{RI$^\prime$}}}_c}{\partial \alpha} 
\end{eqnarray} 
though $Z^{\mbox{\footnotesize{RI$^\prime$}}}_\alpha$ $=$ $1$ at three loops
implies that 
\begin{equation} 
\gamma^{\mbox{\footnotesize{RI$^\prime$}}}_A(a) ~=~ -~ 
\gamma^{\mbox{\footnotesize{RI$^\prime$}}}_\alpha(a) 
\end{equation} 
to the same order which corresponds to the gluon propagator being transverse.
Hence, from the renormalization constants we have computed we find, in four 
dimensions, that 
\begin{eqnarray} 
\gamma^{\mbox{\footnotesize{RI$^\prime$}}}_A(a) &=&  \left[ 8 T_F \Nf
- ( 13 - 3\alpha) C_A \right] \frac{a}{6} \nonumber \\
&& -~ \left[ \left( 27\alpha^3 - 90\alpha^2 - 426\alpha + 3727 \right) C_A^2 
+ \left( 72 \alpha^2 + 240\alpha - 3616 \right) C_A T_F \Nf \right. 
\nonumber \\
&& \left. ~~~~-~ 864 C_F T_F \Nf + 640 T_F^2 \Nf^2 \right] \frac{a^2}{216} 
\nonumber \\
&& +~ \left[ 51200 T_F^3 \Nf^3 - 15552 C_F^2 T_F \Nf + \left( 331776 \zeta(3)
- 487296 \right) C_F T_F^2 \Nf^2 \right. \nonumber \\
&& \left. ~~~~-~ \left( 486\alpha^5 + 3078\alpha^4 + 10260\alpha^3 
- 1458 \zeta(3) \alpha^2 - 25965\alpha^2 + 86184 \zeta(3) \alpha \right. 
\right. \nonumber \\
&& \left. \left. ~~~~~~~~~-~ 173406 \alpha - 175446 \zeta(3) + 2127823 \right) 
C_A^3 ~-~ \left( 648 \alpha^4 
+ 216\alpha^3 + 47808 \alpha^2 \right. \right. \nonumber \\
&& \left. \left. ~~~~~~~~~+~ 10368 \zeta(3) \alpha + 126480 \alpha 
- 254016 \zeta(3) - 2501184 \right) C_A^2 T_F \Nf \right. \nonumber \\
&& \left. ~~~~-~ \left( 7776 \alpha^2 - 62208 \zeta(3) \alpha + 71280 \alpha 
+ 725760 \zeta(3) - 1131408 \right) C_A C_F T_F \Nf \right. \nonumber \\
&& \left. ~~~~+~ \left( 11520 \alpha^2 + 19200 \alpha - 165888 \zeta(3) 
- 751680 \right) C_A T_F^2 \Nf^2 \right] \frac{a^3}{7776} ~+~ O(a^4) 
\nonumber \\ 
\end{eqnarray} 
\begin{eqnarray} 
\gamma^{\mbox{\footnotesize{RI$^\prime$}}}_\psi(a) &=&  \alpha C_F a ~+~ 
\left[ \left( 9 \alpha^3 + 45 \alpha^2 + 223 \alpha + 225 \right) C_A
- 54 C_F - \left( 80 \alpha + 72 \right) T_F \Nf \right] \frac{C_F a^2}{36} 
\nonumber \\
&& +~ \left[ \left( 162 \alpha^5 + 1377 \alpha^4 + 7578 \alpha^3 
+ 1134 \zeta(3) \alpha^2 + 22608 \alpha^2 \right. \right. \nonumber \\
&& \left. \left. ~~~~~-~ 23004 \zeta(3) \alpha + 113080 \alpha 
- 39690 \zeta(3) + 179811 \right) C_A^2 \right. \nonumber \\
&& \left. ~~~~~-\, \left( 648 \alpha^3 + 1944 \alpha^2 - 15552 \zeta(3) 
+ 52272 \right) C_A C_F \right. \nonumber \\
&& \left. ~~~~~-\, \left( 1440 \alpha^3 + 7200 \alpha^2 + 5184 \zeta(3) 
\alpha + 63616 \alpha - 10368 \zeta(3) + 110016 \right) C_A T_F \Nf \right.
\nonumber \\
&& \left. ~~~~~+~ \left( 20736 \zeta(3) \alpha - 23760 \alpha + 6048 \right) 
C_F T_F \Nf \right. \nonumber \\
&& \left. ~~~~~+~ \left( 6400 \alpha + 14976 \right) T_F^2 \Nf^2 + 1944 C_F^2 
\right] \frac{C_F a^3}{1296} ~+~ O(a^4) 
\end{eqnarray} 
and 
\begin{eqnarray} 
\gamma^{\mbox{\footnotesize{RI$^\prime$}}}_c(a) &=&  \left[ \alpha - 3 \right]
\frac{C_A a}{4} ~+~ \left[ \left( 9 \alpha^3 + 18 \alpha^2 + 88 \alpha - 813
\right) C_A - \left( 80 \alpha - 312 \right) T_F \Nf \right] 
\frac{C_A a^2}{144} \nonumber \\
&& +~ \left[ \left( 162 \alpha^5 + 891 \alpha^4 + 972 \zeta(3) \alpha^3 
+ 1503 \alpha^3 + 4050 \zeta(3) \alpha^2 - 2070 \alpha^2 \right. \right.
\nonumber \\
&& \left. \left. ~~~~~-~ 15876 \zeta(3) \alpha + 46363 \alpha + 34182 \zeta(3) 
- 471909 \right) C_A^2 \right. \nonumber \\
&& \left. ~~~~-~ \left( 1440 \alpha^3 + 2880 \alpha^2 + 7776 \zeta(3) \alpha 
+ 39208 \alpha - 33696 \zeta(3) - 322680 \right) C_A T_F \Nf \right. 
\nonumber \\
&& \left. ~~~~+~ \left( 20736 \zeta(3) \alpha - 23760 \alpha - 62208 \zeta(3) 
+ 79056 \right) C_F T_F \Nf \right. \nonumber \\
&& \left. ~~~~+~ \left( 6400 \alpha - 48000 \right) T_F^2 \Nf^2 \right] 
\frac{C_A a^3}{5184} ~+~ O(a^4) 
\end{eqnarray} 
where we use the same convention for the renormalization group functions as for
the renormalization constants in that the variables $a$ and $\alpha$ are in the
scheme indicated on the renormalization group function itself. The expression 
for $\gamma^{\mbox{\footnotesize{RI$^\prime$}}}_\psi(a)$ agrees with the three
loop expression for the colour group $SU(N_c)$ in the Landau gauge given in 
\cite{12}. Moreover, a final check on our calculation resides in the fact that 
in constructing these RI$^\prime$ scheme renormalization group functions the 
correct double and triple poles in $\epsilon$ in the renormalization constants 
have been determined in the computation. If they were not correct then 
{\em finite} renormalization group functions would {\em not} have emerged. 

In \cite{12} the quark anomalous dimension was computed explicitly by first 
determining the appropriate function which converts the $\MSbar$ result to the 
RI$^\prime$ expression following a standard procedure which is discussed in, 
for example, \cite{31}. As a final check on our wave function renormalization
group functions in the RI$^\prime$ scheme we have also computed them from the 
conversion functions which are defined as 
\begin{equation} 
C_A(a,\alpha) ~=~ \frac{Z^{\mbox{\footnotesize{RI$^\prime$}}}_A}  
{Z^{\mbox{\footnotesize{$\MSbar$}}}_A} ~~~,~~~  
C_c(a,\alpha) ~=~ \frac{Z^{\mbox{\footnotesize{RI$^\prime$}}}_c}  
{Z^{\mbox{\footnotesize{$\MSbar$}}}_c} ~~~,~~~  
C_\psi(a,\alpha) ~=~ \frac{Z^{\mbox{\footnotesize{RI$^\prime$}}}_\psi}  
{Z^{\mbox{\footnotesize{$\MSbar$}}}_\psi} ~. 
\end{equation} 
It is important to appreciate how these functions are explicitly constructed.
They are functions of the two parameters $a$ and $\alpha$ in the same scheme.
However, the RI$^\prime$ scheme renormalization constants depend on 
$a_{\mbox{\footnotesize{RI$^\prime$}}}$ and  
$\alpha_{\mbox{\footnotesize{RI$^\prime$}}}$ which therefore must be converted
to their $\MSbar$ counterparts. In the following expressions for the conversion
functions, and those we give later, we have chosen to express them in terms of 
the $\MSbar$ variables and omitted the corresponding subscript. We found 
\begin{eqnarray}  
C_A(a,\alpha) &=& 1 ~+~ \left[ \left( 9 \alpha^2 + 18 \alpha + 97 \right) C_A
- 80 T_F \Nf \right] \frac{a}{36} \nonumber \\ 
&& +~ \left[ \left( 810 \alpha^3 + 2430 \alpha^2 + 5184 \zeta(3) \alpha 
+ 2817 \alpha - 7776 \zeta(3) + 83105 \right) C_A^2 \right. \nonumber \\
&& \left. ~~~~~~-~ \left( 2880 \alpha + 20736 \zeta(3) + 69272 \right) 
C_A T_F \Nf + \left( 41472 \zeta(3) - 47520 \right) C_F T_F \Nf \right.
\nonumber \\
&& \left. ~~~~~~+~ 12800 T_F^2 \Nf^2 \right] \frac{a^2}{2592} \nonumber \\
&& +~ \left[ \left( 17010 \zeta(5) \alpha^4 - 12636 \zeta(3) \alpha^4
+ 64638 \alpha^4 - 51516 \zeta(3) \alpha^3 + 19440 \zeta(5) \alpha^3 
\right. \right. \nonumber \\
&& \left. \left. ~~~~~~+~ 322947 \alpha^3 + 203148 \zeta(3) \alpha^2 
- 8748 \zeta(4) \alpha^2 - 374220 \zeta(5) \alpha^2 + 1094553 \alpha^2 
\right. \right. \nonumber \\
&& \left. \left. ~~~~~~+~ 4636764 \zeta(3) \alpha - 34992 \zeta(4) \alpha 
- 1341360 \zeta(5) \alpha + 1457685 \alpha 
\right. \right. \nonumber \\
&& \left. \left. ~~~~~~-~ 7531056 \zeta(3) - 26244 \zeta(4) - 3414150 \zeta(5) 
+ 44961125 \right) C_A^3 \right. \nonumber \\
&& \left. ~~~~+~ \left( 15552 \zeta(3) \alpha^2 - 303912 \alpha^2 
- 3670272 \zeta(3) \alpha - 890064 \alpha - 7293888 \zeta(3) 
\right. \right. \nonumber \\
&& \left. \left. ~~~~~~~~~+~ 839808 \zeta(4) + 4976640 \zeta(5)
- 49928712 \right) C_A^2 T_F \Nf \right. \nonumber \\
&& \left. ~~~~+~ \left( 746496 \zeta(3) \alpha - 855360 \alpha 
+ 23514624 \zeta(3) - 1119744 \zeta(4) 
\right. \right. \nonumber \\
&& \left. \left. ~~~~~~~~~+~ 7464960 \zeta(5) - 37099872 \right) C_A C_F T_F 
\Nf \right. \nonumber \\
&& \left. ~~~~+~ \left( 663552 \zeta(3) \alpha + 64512 \alpha 
+ 5971968 \zeta(3) + 13873536 \right) C_A T_F^2 \Nf^2 \right. \nonumber \\
&& \left. ~~~~+~ \left( 9206784 \zeta(3) - 14929920 \zeta(5) + 2965248 \right) 
C_F^2 T_F \Nf \right. \nonumber \\
&& \left. ~~~~+~ \left( 16130304 - 12939264 \zeta(3) \right) C_F T_F^2 \Nf^2
\right. \nonumber \\
&& \left. ~~~~-~ 1024000 T_F^3 \Nf^3 \right] \frac{a^3}{93312} ~+~ O(a^4)  
\end{eqnarray} 
\begin{eqnarray}  
C_c(a,\alpha) &=& 1 ~+~ C_A a \nonumber \\
&& +~ \left[ \left( 72 \alpha^2 - 36 \zeta(3) \alpha^2 + 72 \zeta(3) \alpha
- 21 \alpha - 180 \zeta(3) + 1943 \right) C_A - 760 T_F \Nf \right] 
\frac{C_A a^2}{192} \nonumber \\
&& +~ \left[ \left( 29241 \alpha^3 - 11178 \zeta(3) \alpha^3 - 4860 \zeta(5)
\alpha^3 - 56862 \zeta(3) \alpha^2 + 1458 \zeta(4) \alpha^2 \right. \right.
\nonumber \\
&& \left. \left. ~~~~~+~ 34020 \zeta(5) \alpha + 102789 \alpha^2 
+ 254826 \zeta(3) \alpha + 5832 \zeta(4) \alpha - 63180 \zeta(5) \alpha
\right. \right. \nonumber \\
&& \left. \left. ~~~~~-~ 3510 \alpha - 728082 \zeta(3) + 4374 \zeta(4)
- 63180 \zeta(5) + 4329412 \right) C_A^2 \right. \nonumber \\
&& \left. ~~~~+~ \left( 42984 \alpha - 67392 \zeta(3) \alpha - 100224 \zeta(3) 
- 139968 \zeta(4) - 2650192 \right) C_A T_F \Nf \right. \nonumber \\
&& \left. ~~~~+~ \left( 684288 \zeta(3) + 186624 \zeta(4) - 1165104 \right) 
C_F T_F \Nf \right. \nonumber \\
&& \left. ~~~~+~ \left( 27648 \zeta(3) + 330304 \right) T_F^2 \Nf^2 \right] 
\frac{C_A a^3}{31104} ~+~ O(a^4) 
\end{eqnarray} 
\begin{eqnarray}  
C_\psi(a,\alpha) &=& 1 ~-~ \alpha C_F a \nonumber \\
&& +~ \left[ \left( 8 \alpha^2 + 5 \right) C_F - \left( 9 \alpha^2 
- 24 \zeta(3) \alpha + 52 \alpha - 24 \zeta(3) + 82 \right) C_A 
+ 28 T_F \Nf \right] \frac{C_F a^2}{8} \nonumber \\ 
&& +~ \left[ \left( 1728 \zeta(3) \alpha^3 - 11880 \alpha^3 + 25272 \zeta(3)
\alpha^2 - 972 \zeta(4) \alpha^2 - 6480 \zeta(5) \alpha^2 \right. \right.
\nonumber \\
&& \left. \left. ~~~~~-~ 63747 \alpha^2 + 181440 \zeta(3) \alpha 
- 1944 \zeta(4) \alpha - 12960 \zeta(5) \alpha - 358191 \alpha \right. \right.
\nonumber \\
&& \left. \left. ~~~~~+~ 678024 \zeta(3) + 22356 \zeta(4) - 213840 \zeta(5) 
- 1274056 \right) C_A^2 \right. \nonumber \\
&& \left. ~~~~+~ \left( 12312 \alpha^2 - 5184 \zeta(3) \alpha^3 
- 31104 \zeta(3) \alpha^2 + 59616 \alpha^2 + 57024 \zeta(3) \alpha 
\right. \right. \nonumber \\
&& \left. \left. ~~~~~~~~~-~ 103680 \zeta(5) \alpha + 85536 \alpha 
- 228096 \zeta(3) - 31104 \zeta(4) \right. \right. \nonumber \\
&& \left. \left. ~~~~~~~~~+~ 103680 \zeta(5) + 215352 \right) C_A C_F
\right. \nonumber \\
&& \left. ~~~~+~ \left( 3456 \zeta(3) \alpha^3 - 5184 \alpha^3 - 11016 \alpha 
+ 31536 \right) C_F^2 \right. \nonumber \\
&& \left. ~~~~+~ \left( 124056 \alpha - 41472 \zeta(3) \alpha - 89856 \zeta(3) 
+ 760768 \right) C_A T_F \Nf \right. \nonumber \\
&& \left. ~~~~+~ \left( 68256 - 82944 \zeta(3) - 28512 \alpha \right) C_F T_F 
\Nf \right. \nonumber \\
&& \left. ~~~~-~ 100480 T_F^2 \Nf^2 \right] \frac{C_F a^3}{5184} ~+~ O(a^4) ~. 
\end{eqnarray} 
With these the RI$^\prime$ scheme renormalization group functions can be
determined from 
\begin{eqnarray}
\gamma^{\mbox{\footnotesize{RI$^\prime$}}}_i 
\left(a_{\mbox{\footnotesize{RI$^\prime$}}}\right) &=& 
\gamma^{\mbox{\footnotesize{$\MSbar$}}}_i 
\left(a_{\mbox{\footnotesize{$\MSbar$}}}\right) ~+~ 
\beta\left(a_{\mbox{\footnotesize{$\MSbar$}}}\right) 
\frac{\partial ~}{\partial a_{\mbox{\footnotesize{$\MSbar$}}}} 
\ln C_i\left(a_{\mbox{\footnotesize{$\MSbar$}}}, 
\alpha_{\mbox{\footnotesize{$\MSbar$}}}\right) \nonumber \\
&& +~ \alpha_{\mbox{\footnotesize{$\MSbar$}}} 
\gamma^{\mbox{\footnotesize{$\MSbar$}}}_\alpha 
\left(a_{\mbox{\footnotesize{$\MSbar$}}}\right) 
\frac{\partial ~}{\partial \alpha_{\mbox{\footnotesize{$\MSbar$}}}}  
\ln C_i\left(a_{\mbox{\footnotesize{$\MSbar$}}},  
\alpha_{\mbox{\footnotesize{$\MSbar$}}}\right) 
\label{gammari}
\end{eqnarray}
where $i$ $=$ $A$, $c$ or $\psi$ and we have included the scheme dependence of
the variables explicitly though to the order we are working to the 
$\beta$-function is the same in both schemes. For each of the three cases we 
have computed the right hand side of (\ref{gammari}) in terms of the $\MSbar$ 
variables and then converted to their RI$^\prime$ counterparts before verifying
that the same previous expressions correctly emerge in terms of the 
RI$^\prime$ scheme variables. This completes the full three loop 
renormalization of the QCD Lagrangian in the RI$^\prime$ scheme. 

\sect{Quark mass anomalous dimension in the RI$^\prime$ scheme.} 

We now turn to the problem of deducing similar renormalization constants for
the composite quark currents of the form ${\cal O}_{\cal A}$~$=$~$\bar{\psi} 
{\cal A} \psi$ where ${\cal A}$ $=$ $1$, $\gamma^\mu$ or $\sigma^{\mu\nu}$ with
$\sigma^{\mu\nu}$ $=$ $\half [ \gamma^\mu, \gamma^\nu ]$ where the latter 
corresponds to the tensor current. As the lattice data is available for the 
insertion of the operator at zero momentum to deduce the conversion functions 
we will insert each of ${\cal O}_{\cal A}$ at zero momentum into a quark 
two-point function and examine the divergence structure of 
\begin{equation} 
G_{{\cal O}_{\cal A}}(p) ~=~ \langle \psi(p) ~ [ \bar{\psi} {\cal A} 
\psi ](0) ~ \bar{\psi}(-p) \rangle ~=~ \langle \psi(p) ~ {\cal O}_{\cal A}(0) ~
\bar{\psi}(-p) \rangle ~. 
\end{equation}
In order to demonstrate the validity of this approach we must first reconstruct
the anomalous dimension for the mass, \cite{12}, which corresponds to the 
operator ${\cal A}$~$=$~$1$. Therefore, defining the renormalization constant 
$Z_{\bar{\psi} \psi}$ of the operator in the usual way by,  
\begin{equation} 
{\cal O}_{1\,\mbox{\footnotesize{o}}} ~=~ Z_{\bar{\psi}\psi} {\cal O}_1 
\end{equation} 
the RI$^\prime$ scheme value is given by the condition 
\begin{equation} 
\left. \lim_{\epsilon \rightarrow 0} \left[ 
Z^{\mbox{\footnotesize{RI$^\prime$}}}_{\bar{\psi}\psi}  
Z^{\mbox{\footnotesize{RI$^\prime$}}}_\psi 
\langle \psi(p) {\cal O}_{\bar{\psi}\psi} (0) \bar{\psi}(-p) \rangle \right] 
\right|_{p^2 \, = \, \mu^2} ~=~ 1  
\end{equation}  
where the wave function renormalization constants arise from the external
fields. In the $\MSbar$ scheme this results in a gauge independent 
renormalization constant to all orders, \cite{1,3}. However, in RI$^\prime$ 
this is not the case since
\begin{eqnarray}
Z^{\mbox{\footnotesize{RI$^\prime$}}}_{\bar{\psi} \psi} &=& 1 ~-~ \left(
\frac{3}{\epsilon} + 4 + \alpha \right) C_F a ~+~ \left[ \left( \frac{9}{2} 
C_F^2 + \frac{11}{2} C_F C_A - 2 T_F \Nf C_F \right) \frac{1}{\epsilon^2} 
\right. \nonumber \\
&& \left. +~ \left( \frac{5}{3} T_F \Nf C_F - \frac{97}{12} C_F C_A 
+ \left( \frac{45}{4} + 3 \alpha \right) C_F^2 \right) \frac{1}{\epsilon}
\right. \nonumber \\
&& \left. +~ \left( \left( \frac{83}{6} + \frac{20\alpha}{9} \right) 
T_F \Nf C_F + \left( 18 \zeta(3) - \frac{1285}{24} - \frac{223\alpha}{36} 
- \frac{5\alpha^2}{4} - \frac{\alpha^3}{4} \right) C_F C_A \right. \right. 
\nonumber \\
&& \left. \left. ~~~~~+~ \left( \frac{19}{8} - 12 \zeta(3) + 4 \alpha 
+ \alpha^2 \right) C_F^2 \right) \right] a^2 \nonumber \\
&& +~ \left[ \left( \frac{88}{9} T_F \Nf C_F C_A + 6 T_F \Nf C_F^2 
- \frac{16}{9} T_F^2 \Nf^2 C_F - \frac{121}{9} C_F C_A^2 - \frac{33}{2} C_F^2
C_A - \frac{9}{2} C_F^3 \right) \frac{1}{\epsilon^3} \right. \nonumber \\
&& \left. ~~~~~+~ \left( \frac{40}{27} T_F^2 \Nf^2 C_F - \frac{484}{27} 
T_F C_F C_A + \left( 2 \alpha - \frac{5}{3} \right) T_F \Nf C_F^2 
+ \frac{1679}{54} C_F C_A^2 \right. \right. \nonumber \\
&& \left. \left. ~~~~~+~ \left( \frac{49}{12} - \frac{11\alpha}{2} \right) 
C_F^2 C_A - \left( \frac{63}{4} + \frac{9\alpha}{2} \right) C_F^3 \right) 
\frac{1}{\epsilon^2} \right. \nonumber \\
&& \left. +~ \left( \left( \frac{556}{81} + 16 \zeta(3) \right) T_F \Nf C_F C_A
- \left( \frac{197}{6} + 16 \zeta(3) + \frac{25\alpha}{3} \right) T_F \Nf C_F^2
+ \frac{140}{81} T_F^2 \Nf^2 C_F \right. \right. \nonumber \\
&& \left. \left. ~~~~~~-~ \frac{11413}{324} C_F C_A^2 + \left( \frac{4889}{24} 
- 54 \zeta(3) + \frac{80\alpha}{3} + \frac{15\alpha^2}{4} + \frac{3\alpha^3}{4}
\right) C_F^2 C_A \right. \right. \nonumber \\
&& \left. \left. ~~~~~~+~ \left( 36 \zeta(3) - \frac{205}{8} 
- \frac{45\alpha}{4} - 3 \alpha^2 \right) C_F^3 \right) \frac{1}{\epsilon} 
\right. \nonumber \\ 
&& \left. -~ \left( \left( \frac{616}{9} \zeta(3) - 4 \alpha \zeta(3) 
- \frac{95387}{243} - 24 \zeta(4) - \frac{3976\alpha}{81} 
- \frac{50\alpha^2}{9} - \frac{10\alpha^3}{9} \right) T_F \Nf C_F C_A \right. 
\right. \nonumber \\
&& \left. \left. ~~~~~+~ \left( \frac{128}{3} \zeta(3) + 8 \alpha \zeta(3) 
+ 24 \zeta(4) - \frac{1109}{9} + \frac{115\alpha}{18} + \frac{40\alpha^2}{9} 
\right) T_F \Nf C_F^2 \right. \right. \nonumber \\
&& \left. \left. ~~~~~+~ \left( \frac{7514}{243} + \frac{400\alpha}{81} 
+ \frac{32}{9} \zeta(3) \right) T_F^2 \Nf^2 C_F 
+ \left( \frac{3360023}{3888} + \frac{28351\alpha}{324} 
+ \frac{157\alpha^2}{9} \right. \right. \right. \nonumber \\
&& \left. \left. \left. ~~~~~~~~+~ \frac{421\alpha^3}{72} 
+ \frac{17\alpha^4}{16} + \frac{\alpha^5}{8} - \frac{31193}{72} \zeta(3) 
- \frac{65\alpha}{4} \zeta(3) + \frac{7\alpha^2}{8} \zeta(3) + 60 \zeta(5) 
\right) C_F C_A^2 \right. \right. \nonumber \\
&& \left. \left. ~~~~~+~ \left( \left( \frac{962}{3} + 43 \alpha + 3 \alpha^2 
\right) \zeta(3) - 20 \zeta(5) - \frac{18781}{72} - \frac{6089\alpha}{72} 
\right. \right. \right. \nonumber \\
&& \left. \left. \left. ~~~~~~~~~~~-~ \frac{170\alpha^2}{9} - 4 \alpha^3 
- \frac{\alpha^4}{2} \right) C_F^2 C_A + \left( \frac{3227}{12} 
+ \frac{31\alpha}{8} + 4 \alpha^2 + \alpha^3 \right. \right. \right. 
\nonumber \\
&& \left. \left. \left. ~~~~~~~~~~~-~ \left( 58 - 6 \alpha + 6 \alpha^2 \right) 
\zeta(3) - 120 \zeta(5) \right) C_F^3 \frac{}{} \right) \right] a^3 ~+~ O(a^4) 
\end{eqnarray}  
which agrees with the quark mass renormalization constant 
$Z^{\mbox{\footnotesize{RI$^\prime$}}}_m$ deduced at four loops in \cite{12} in
the Landau gauge in our conventions. Thus we have demonstrated the equivalence 
of our operator method with the massive propagator approach of \cite{12}.
Further, we have checked that the correct three loop $\MSbar$ quark mass
anomalous dimension, \cite{32,33}, emerges from our programmes. Therefore, from 
$Z^{\mbox{\footnotesize{RI$^\prime$}}}_m$ we can deduce the corresponding 
renormalization group function. With  
\begin{equation}
\gamma^{\mbox{\footnotesize{RI$^\prime$}}}_{\bar{\psi}\psi}(a) ~=~ -~ \beta(a) 
\frac{\partial \ln Z^{\mbox{\footnotesize{RI$^\prime$}}}_{\bar{\psi}\psi}} 
{\partial a} ~-~ \alpha \gamma^{\mbox{\footnotesize{RI$^\prime$}}}_\alpha(a) 
\frac{\partial \ln Z^{\mbox{\footnotesize{RI$^\prime$}}}_{\bar{\psi}\psi}} 
{\partial \alpha} 
\end{equation}
we find 
\begin{eqnarray} 
\gamma^{\mbox{\footnotesize{RI$^\prime$}}}_{\bar{\psi}\psi}(a) &=& -~ 
3 C_F a ~-~ [ ( 185 + 9 \alpha + 3 \alpha^2 ) C_A ~+~ 9C_F ~-~ 52 T_F \Nf ] 
\frac{C_F a^2}{6} \nonumber \\ 
&& +~ \left[ \left( 108 \alpha^3 + 324 \alpha^2 - 1944 - 19008 \zeta(3) \right) 
C_A C_F \right. \nonumber \\ 
&& \left. ~~~~-~ \left( 117428 + 5634 \alpha + 1905 \alpha^2 + 405 \alpha^3 
+ 54 \alpha^4 - 28512 \zeta(3) \right) C_A^2 \right. \nonumber \\
&& \left. ~~~~+~ \left( 480 \alpha^2 + 2088 \alpha + 62960 \right) 
C_A T_F \Nf ~-~ 13932 C_F^2 \right. \nonumber \\
&& \left. ~~~~+~ \left( 16632 - 3456 \zeta(3) \right) C_F T_F \Nf ~-~ 
6848 T_F^2 \Nf^2) \right] \frac{C_F a^3}{216} ~+~ O(a^4)  
\label{gammamris} 
\end{eqnarray}  
which agrees with \cite{12} in the Landau gauge, aside from an overall factor 
stemming from our conventions which are the same as \cite{34}. We have also 
derived the same expression by constructing the conversion function 
$C_{{\cal O}_{\cal A}}(a,\alpha)$ with ${\cal A}$ $=$ $1$ where 
\begin{equation} 
C_{{\cal O}_{\cal A}}(a,\alpha) ~=~ 
\frac{Z^{\mbox{\footnotesize{RI$^\prime$}}}_{{\cal O}_{\cal A}}} 
{Z^{\mbox{\footnotesize{$\MSbar$}}}_{{\cal O}_{\cal A}}} ~. 
\end{equation} 
Thus, using 
\begin{eqnarray}
\gamma^{\mbox{\footnotesize{RI$^\prime$}}}_{{\cal O}_{\cal A}} 
\left(a_{\mbox{\footnotesize{RI$^\prime$}}}\right) &=& 
\gamma^{\mbox{\footnotesize{$\MSbar$}}}_{{\cal O}_{\cal A}} 
\left(a_{\mbox{\footnotesize{$\MSbar$}}}\right) ~-~ 
\beta\left(a_{\mbox{\footnotesize{$\MSbar$}}}\right) 
\frac{\partial ~}{\partial a_{\mbox{\footnotesize{$\MSbar$}}}} 
\ln C_{{\cal O}_{\cal A}}\left(a_{\mbox{\footnotesize{$\MSbar$}}}, 
\alpha_{\mbox{\footnotesize{$\MSbar$}}}\right) \nonumber \\
&& -~ \alpha_{\mbox{\footnotesize{$\MSbar$}}} 
\gamma^{\mbox{\footnotesize{$\MSbar$}}}_\alpha 
\left(a_{\mbox{\footnotesize{$\MSbar$}}}\right) 
\frac{\partial ~}{\partial \alpha_{\mbox{\footnotesize{$\MSbar$}}}}  
\ln C_{{\cal O}_{\cal A}}\left(a_{\mbox{\footnotesize{$\MSbar$}}},  
\alpha_{\mbox{\footnotesize{$\MSbar$}}}\right) 
\end{eqnarray} 
with the explicit expression 
\begin{eqnarray} 
C_{\bar{\psi}\psi}(a,\alpha) &=& 1 ~-~ (\alpha + 4) C_F a ~+~ \left[ \left( 
24 \alpha^2 + 96 \alpha - 288 \zeta(3) + 57 \right) C_F \right. \nonumber \\
&& \left. ~~+~ 332 T_F \Nf - \left( 18 \alpha^2 + 84\alpha - 432 \zeta(3) 
+ 1285 \right) C_A \right] \frac{C_F a^2}{24} \nonumber \\
&& +~ \left[ \left( 15552 \alpha^3 + 89424 \alpha^2 - 23328 \alpha^2 \zeta(3) 
+ 573804 \alpha \right. \right. \nonumber \\
&& \left. \left. ~~~~~~-~ 334368 \alpha \zeta(3) - 2493504 \zeta(3) 
+ 155520 \zeta(5) + 2028348 \right) C_A C_F \right. \nonumber \\
&& \left. ~~~~-~ \left( 13122 \alpha^3 - 8748 \alpha^2 \zeta(3) 
+ 71685 \alpha^2 - 103032 \alpha \zeta(3) \right. \right. \nonumber \\
&& \left. \left. ~~~~~~~~~~+~ 357777 \alpha - 3368844 \zeta(3) 
+ 466560 \zeta(5) + 6720046 \right) C_A^2 \right. \nonumber \\
&& \left. ~~~~+~ \left( 113400 \alpha - 31104 \zeta(3) \alpha \right. \right.
\nonumber \\
&& \left. \left. ~~~~~~~~~~-~ 532224 \zeta(3) + 186624 \zeta(4) + 3052384 
\right) C_A T_F \Nf \right. \nonumber \\
&& \left. ~~~~+~ \left( 62208 \alpha \zeta(3) - 123120 \alpha - 331776 \zeta(3)
\right. \right. \nonumber \\
&& \left. \left. ~~~~~~~~~~-~ 186624 \zeta(4) + 958176 \right) C_F T_F \Nf 
\right. \nonumber \\
&& \left. ~~~~-~ \left( 7776 \alpha^3 - 46656 \alpha^2 \zeta(3) 
+ 31104 \alpha^2 + 46656 \alpha \zeta(3) + 30132 \alpha \right. \right. 
\nonumber \\
&& \left. \left. ~~~~~~~~~~-~ 451008 \zeta(3) - 933120 \zeta(5) + 2091096 
\right) C_F^2 \right. \nonumber \\
&& \left. ~~~~-~ \left( 27648 \zeta(3) + 240448 \right) T_F^2 \Nf^2 \right] 
\frac{C_F a^3}{7776} ~+~ O(a^4)  
\end{eqnarray}  
we find exact agreement. Moreover, this conversion function agrees with that
given in \cite{12} when restricted to the Landau gauge. However, in checking
the three loop expression in the RI$^\prime$ scheme only the contribution up to
and including the two loop term is required. The three loop part is only 
relevant for the four loop anomalous dimension. Therefore, since the four loop
$\MSbar$ quark mass anomalous dimension is available, \cite{5,6}, for an
arbitrary colour group we can deduce that the four loop correction to  
(\ref{gammamris}) is 
\begin{eqnarray} 
\gamma^{\mbox{\footnotesize{RI$^\prime$}}}_{\bar{\psi}\psi}(a) &=& -~ 
3 C_F a ~-~ [ ( 185 + 9 \alpha + 3 \alpha^2 ) C_A ~+~ 9C_F ~-~ 52 T_F \Nf ] 
\frac{C_F a^2}{6} \nonumber \\ 
&& +~ \left[ \left( 108 \alpha^3 + 324 \alpha^2 - 1944 - 19008 \zeta(3) \right) 
C_A C_F \right. \nonumber \\ 
&& \left. ~~~~-~ \left( 117428 + 5634 \alpha + 1905 \alpha^2 + 405 \alpha^3 
+ 54 \alpha^4 - 28512 \zeta(3) \right) C_A^2 \right. \nonumber \\
&& \left. ~~~~+~ \left( 480 \alpha^2 + 2088 \alpha + 62960 \right) 
C_A T_F \Nf ~-~ 13932 C_F^2 \right. \nonumber \\
&& \left. ~~~~+~ \left( 16632 - 3456 \zeta(3) \right) C_F T_F \Nf ~-~ 
6848 T_F^2 \Nf^2) \right] \frac{C_F a^3}{216} \nonumber \\
&& +~ \left[ \left( - 1215 \alpha^6 - 13608 \alpha^5 - 90801 \alpha^4 
- 8262 \zeta(3) \alpha^3 - 368064 \alpha^3 + 104004 \zeta(3) \alpha^2
\right. \right. \nonumber \\
&& \left. \left. ~~~~~~-~ 1397826 \alpha^2 + 940734 \zeta(3) \alpha 
- 4554684 \alpha + 39004740 \zeta(3) \right. \right. \nonumber \\
&& \left. \left. ~~~~~~-~ 1710720 \zeta(5) - 92569118 \right) C_A^3 C_F 
\right. \nonumber \\
&& \left. ~~~~~~+~ \left( 2916 \alpha^5 + 28188 \alpha^4 - 23328 \zeta(3) 
\alpha^3 + 136404 \alpha^3 - 252720 \zeta(3) \alpha^2 \right. \right. 
\nonumber \\
&& \left. \left. ~~~~~~~~~~~~+~ 377136 \alpha^2 - 1717200 \zeta(3) \alpha
+ 429300 \alpha - 22203072 \zeta(3) \right. \right. \nonumber \\
&& \left. \left. ~~~~~~~~~~~~-~ 1710720 \zeta(5) + 10355148 \right) C_A^2 C_F^2 
\right. \nonumber \\
&& \left. ~~~~~~+~ \left( 12960 \alpha^4 + 103032 \alpha^3 + 34992 \zeta(3) 
\alpha^2 \right. \right. \nonumber \\
&& \left. \left. ~~~~~~~~~~~~+~ 677952 \alpha^2 - 316224 \zeta(3) \alpha 
+ 3021840 \alpha - 14239152 \zeta(3) \right. \right. \nonumber \\
&& \left. \left. ~~~~~~~~~~~~-~ 1244160 \zeta(5) + 73217928 \right) 
C_A^2 C_F T_F \Nf \right. \nonumber \\
&& \left. ~~~~~~+~ \left( -~ 3888 \alpha^4 + 46656 \zeta(3) \alpha^3 
- 11664 \alpha^3 + 241056 \zeta(3) \alpha^2 - 5832 \alpha^2 \right. \right.
\nonumber \\
&& \left. \left. ~~~~~~~~~~~~-~ 1236384 \zeta(3) \alpha - 103032 \alpha 
- 1601856 \zeta(3) \right. \right. \nonumber \\
&& \left. \left. ~~~~~~~~~~~~+~ 10264320 \zeta(5) - 33960384  \right) C_A C_F^3 
\right. \nonumber \\
&& \left. ~~~~~~+~ \left( -~ 25920 \alpha^3 - 62208 \zeta(3) \alpha^2 
+ 18144 \alpha^2 + 694656 \zeta(3) \alpha + 230688 \alpha \right. \right.
\nonumber \\
&& \left. \left. ~~~~~~~~~~~~+~ 1347840 \zeta(3) - 1244160 \zeta(5)
+ 20983248 \right) C_A C_F^2 T_F \Nf \right. \nonumber \\
&& \left. ~~~~~~+~ \left( -~ 57600 \alpha^2 + 82944 \zeta(3) \alpha 
- 449280 \alpha \right. \right. \nonumber \\
&& \left. \left. ~~~~~~~~~~~~+~ 580608 \zeta(3) - 16599552 \right) C_A C_F 
T_F^2 \Nf^2 \right. \nonumber \\
&& \left. ~~~~~~+~ \left( -~ 62208 \alpha^2 + 373248 \zeta(3) \alpha 
+ 31104 \alpha \right. \right. \nonumber \\
&& \left. \left. ~~~~~~~~~~~~-~ 3856896 \zeta(3) + 9745920 \right) C_F^3 T_F 
\Nf \right. \nonumber \\
&& \left. ~~~~~~+~ \left( -~ 165888 \zeta(3) \alpha + 41472 \alpha 
+ 2571264 \zeta(3) - 6653952 \right) C_F^2 T_F^2 \Nf^2 \right. \nonumber \\
&& \left. ~~~~~~+~ \left( 2612736 \zeta(3) + 1225692 \right) C_F^4 
+ 1025536 C_F T_F^3 \Nf^3 \right. \nonumber \\
&& \left. ~~~~~~+~ \left( 248832 - 1866240 \zeta(3) \right) 
\frac{d_A^{abcd}d_F^{abcd}}{N_F} \right. \nonumber \\
&& \left. ~~~~~~+~ \left( 3732480 \zeta(3) - 497664 \right) 
\Nf \frac{d_F^{abcd}d_F^{abcd}}{N_F} \right] \frac{a^4}{7776} ~+~ O(a^5) 
\end{eqnarray}  
where $\frac{d_A^{abcd}d_F^{abcd}}{N_F}$ and $\frac{d_F^{abcd}d_F^{abcd}}{N_F}$
are the quartic Casimirs associated with light-by-light topologies, \cite{4,6},
and the coupling constant and gauge parameter are in the RI$^\prime$ scheme. 
The four loop expression is in agreement with the Landau gauge expression of 
\cite{12} for an $SU(N_c)$ colour group.  

As an additional check on the renormalization of a composite operator in the
RI$^\prime$ scheme we have also considered the case of ${\cal A}$ $=$ 
$\gamma^\mu$. One reason for examining this operator arises from the fact that 
as it is now a Lorentz vector the Green's function, 
$G^\mu_{\bar{\psi}\gamma^\mu\psi}(p)$, is not merely proportional to 
$\gamma^\mu$. Instead by Lorentz symmetry 
\begin{equation} 
G^\mu_{\bar{\psi} \gamma^\mu \psi}(p) ~=~ \langle \psi(p) ~ [ \bar{\psi} 
\gamma^\mu \psi ] (0) ~ \bar{\psi}(-p) \rangle ~=~ \Sigma^{(1)}_{\bar{\psi} 
\gamma^\mu \psi}(p) \gamma^\mu ~+~ \Sigma^{(2)}_{\bar{\psi} \gamma^\mu \psi}(p)
\frac{p^\mu \pslash}{p^2}  
\label{vecdec} 
\end{equation}  
where the amplitudes $\Sigma^{(i)}_{\bar{\psi} \gamma^\mu \psi}(p)$ depend on 
the coupling constant. They are determined by the relations
\begin{eqnarray} 
\Sigma^{(1)}_{\bar{\psi} \gamma^\mu \psi}(p) &=& \frac{1}{4(d-1)} \left[
\mbox{tr} \left( \gamma_\mu G^\mu_{\bar{\psi} \gamma^\mu \psi}(p) \right) ~-~
\mbox{tr} \left( \frac{p_\mu \pslash}{p^2} G^\mu_{\bar{\psi} \gamma^\mu 
\psi}(p) \right) \right] \nonumber \\
\Sigma^{(2)}_{\bar{\psi} \gamma^\mu \psi}(p) &=& -~ \frac{1}{4(d-1)} \left[
\mbox{tr} \left( \gamma_\mu G^\mu_{\bar{\psi} \gamma^\mu \psi}(p) \right) ~-~
d \, \mbox{tr} \left( \frac{p_\mu \pslash}{p^2} G^\mu_{\bar{\psi} \gamma^\mu 
\psi}(p) \right) \right] ~. 
\end{eqnarray}  
The renormalization constant for the operator is determined from
\begin{equation} 
{\cal O}_{\gamma^\mu\,\mbox{\footnotesize{o}}} ~=~ Z_{\bar{\psi}\gamma^\mu\psi}
{\cal O}_{\gamma^\mu} ~.  
\end{equation} 
and renormalizing the Green's function we find 
\begin{equation} 
Z^{\mbox{\footnotesize{$\MSbar$}}}_{\bar{\psi} \gamma^\mu \psi} ~=~ 1 ~+~
O(a^4) ~.  
\end{equation} 
The non-renormalization of this current rests in the fact that it corresponds
to a physical operator and therefore on general grounds its anomalous dimension
vanishes. (See, for example, \cite{31}.) Furthermore, as the vector current has
been inserted at zero momentum the $\gamma^\mu$ component of its Green's 
function must obey the Slavnov-Taylor identity and be equivalent to the finite 
part of the quark two-point function after renormalization in the same scheme. 
In computing the quark wave function anomalous dimension in the previous 
section we have also determined the finite part in the $\MSbar$ scheme and it 
is reassuring to note that both it and 
\begin{eqnarray} 
\left. \Sigma^{(1) ~ \mbox{\footnotesize{$\MSbar$}} ~ 
\mbox{\footnotesize{finite}}}_{\bar{\psi} \gamma^\mu \psi}(p) 
\right|_{p^2 \, = \mu^2} &=& 1 ~+~ \alpha C_F a \nonumber \\
&& +~ \left[ \left( \frac{41}{4} + \frac{13\alpha}{2} + \frac{9\alpha^2}{8} 
- 3 ( 1 + \alpha ) \zeta(3) \right) C_F C_A \right. \nonumber \\ 
&& \left. ~~~~~~~-~ \frac{7}{2} T_F \Nf C_F - \frac{5}{8} C_F^2 \right] a^2 
\nonumber \\
&& +~ \left[ \frac{1570}{81} T_F^2 \Nf^2 + \left( 16 \zeta(3) 
- \frac{79}{6} - \frac{3\alpha}{2} \right) T_F \Nf C_F \right. \nonumber \\
&& \left. ~~~~+~ \left( \frac{52}{3} \zeta(3)
+ 8 \alpha \zeta(3) - \frac{11887}{81} - \frac{1723\alpha}{72} \right) T_F 
\Nf C_A \right. \nonumber \\
&& \left. ~~~~+~ \left( \frac{159257}{648} + \frac{39799\alpha}{576} 
+ \frac{787\alpha^2}{64} + \frac{55\alpha^3}{24} \right. \right. \nonumber \\
&& \left. \left. ~~~~~~~~~~~-~ \left( \frac{3139}{24} + 35 \alpha
+ \frac{39\alpha^2}{8} + \frac{\alpha^3}{3} \right) \zeta(3) \right. \right.
\nonumber \\
&& \left. \left. ~~~~~~~~~~~+~ \left( \frac{3\alpha^2}{16} + \frac{3\alpha}{8} 
- \frac{69}{16} \right) \zeta(4) \right. \right. \nonumber \\
&& \left. \left. ~~~~~~~~~~~+~ \left( \frac{165}{4} + \frac{5\alpha}{2} 
+ \frac{5\alpha^2}{4} \right) \zeta(5) \right) C_A^2 \right. \nonumber \\
&& \left. ~~~~+~ \left( \frac{}{} 20 (\alpha - 1) \zeta(5) + 6 \zeta(4) 
+ \left( 44 - 17 \alpha + \alpha^3 \right) \zeta(3) \right. \right. 
\nonumber \\
&& \left. \left. ~~~~~~~~~~~-~ \frac{997}{24} + 4 \alpha + \frac{3\alpha^2}{2} 
- \frac{\alpha^3}{8} \right) C_F C_A \right. \nonumber \\
&& \left. ~~~~-~ \left( \frac{73}{12} - \frac{7\alpha}{8} + \frac{2\alpha^3}{3} 
\zeta(3) \right) C_F^2 \right] C_F a^3 ~+~ O(a^4) 
\label{vecfinpart}
\end{eqnarray} 
are in exact agreement which provides an additional check on our programming. 
Further, it is the finite part of the first term which determines the relation 
of the $\MSbar$ scheme to the RI$^\prime$ scheme. Such a feature of extra 
contributions will persist in the tensor current case but the vector operator 
is peculiar in the sequence of dimension three operators in that only its 
anomalous dimension {\em and} finite part are entwined with the Slavnov-Taylor 
identity. Repeating the same exercise for the RI$^\prime$ scheme by 
introducing the definition 
\begin{equation} 
\lim_{p^2 \, \rightarrow \, \mu^2} \left( 
Z^{\mbox{\footnotesize{RI$^\prime$}}}_\psi 
Z^{\mbox{\footnotesize{RI$^\prime$}}}_{\bar{\psi} \gamma^\mu \psi} 
\Sigma^{(1)}_{\bar{\psi} \gamma^\mu \psi}(p) \right) ~=~ 1 
\label{vecfinpartris} 
\end{equation} 
we find 
\begin{equation} 
Z^{\mbox{\footnotesize{RI$^\prime$}}}_{\bar{\psi} \gamma^\mu \psi} ~=~ 1 ~+~
O(a^4) ~.  
\end{equation} 
This is consistent with the observation that if an anomalous dimension of a
physical operator vanishes in one scheme it vanishes in any other scheme. 
Moreover, (\ref{vecfinpartris}) is also consistent with the Slavnov-Taylor
identity in the RI$^\prime$ scheme since not only is it finite prior to 
renormalization but it is {\em also} unity which agrees with the finite part 
of the quark two-point function consistent with the nature of this 
renormalization scheme. Finally, to assist with lattice matching we record that
the finite parts of the second component of the vector current Green's function
are 
\begin{eqnarray} 
\left. \Sigma^{(2) ~ \mbox{\footnotesize{$\MSbar$}} ~ 
\mbox{\footnotesize{finite}}}_{\bar{\psi} \gamma^\mu \psi}(p) 
\right|_{p^2 \, = \, \mu^2} &=& -~ 2 \alpha C_F a \nonumber \\
&& -~ C_F \left[ \left( \frac{25}{2} + 7 \alpha + \frac{3}{2} \alpha^2 \right) 
C_A - 4 T_F \Nf + \left( 2 \alpha^2 - 3 \right) C_F \right] \! a^2 \nonumber \\
&& -~ C_F \left[ \left( \frac{28}{3} - 11 \alpha \right) T_F \Nf C_F
+ \frac{208}{9} T_F^2 \Nf^2 + \left( 3 - \frac{17}{4} \alpha \right) C_F^2 
\right. \nonumber \\
&& \left. ~~~~~~~~+~ \left( 16 \zeta(3) + 8 \zeta(3) \alpha - \frac{1528}{9} 
- \frac{175}{6} \alpha \right) T_F \Nf C_A \right. \nonumber \\
&& \left. ~~~~~~~~+~ \left( \frac{19979}{72} + \frac{4393}{48} \alpha 
+ \frac{295}{16} \alpha^2 + \frac{27}{8} \alpha^3 \right. \right. \nonumber \\
&& \left. \left. ~~~~~~~~~~~~~~~~-~ \left( \frac{245}{4} 
+ \frac{59}{2} \alpha + \frac{9}{4} \alpha^2 \right) \zeta(3) \right) C_A^2
\right. \nonumber \\
&& \left. ~~~~~~~~+~ \left( \left( 24 - 6 \alpha - 6 \alpha^2 \right) 
\zeta(3) - \frac{242}{3} + 33 \alpha \right. \right. \nonumber \\
&& \left. \left. ~~~~~~~~~~~~~~~~+~ 17 \alpha^2 + \frac{11}{4} 
\alpha^3 \right) C_F C_A \right] a^3 ~+~ O(a^4) 
\end{eqnarray}  
and 
\begin{eqnarray} 
\left. \Sigma^{(2) ~ \mbox{\footnotesize{RI$^\prime$}} ~
\mbox{\footnotesize{finite}}}_{\bar{\psi} \gamma^\mu \psi}(p) 
\right|_{p^2 \, = \, \mu^2} &=& -~ 2 \alpha C_F a \nonumber \\
&& -~ C_F \left[ \left( \frac{25}{2} + \frac{223}{18} \alpha + \frac{5}{2} 
\alpha^2 + \frac{1}{2} \alpha^3 \right) C_A \right. \nonumber \\
&& \left. ~~~~~~~~-~ \left( 4 + \frac{40}{9} \alpha \right) T_F \Nf 
- 3 C_F \right] a^2 \nonumber \\
&& -~ C_F \left[ \left( \frac{28}{3} - \frac{110}{3} \alpha + 32 \zeta(3) 
\alpha \right) T_F \Nf C_F + \left( \frac{208}{9} + \frac{800}{81} \alpha \! 
\right) T_F^2 \Nf^2 \right. \nonumber \\
&& \left. ~~~~~~~~+~ \left( 16 \zeta(3) - 8 \zeta(3) \alpha - \frac{1528}{9} 
- \frac{7952}{81} \alpha \right. \right. \nonumber \\
&& \left. \left. ~~~~~~~~~~~~~~~~-~ \frac{100}{9} \alpha^2 - \frac{20}{9} 
\alpha^3 \right) T_F \Nf C_A \right. \nonumber \\
&& \left. ~~~~~~~~+~ \left( \frac{19979}{72} + \frac{14135}{81} \alpha 
+ \frac{314}{9} \alpha^2 + \frac{421}{36} \alpha^3 + \frac{17}{8} \alpha^4
+ \frac{1}{4} \alpha^5 \right. \right. \nonumber \\
&& \left. \left. ~~~~~~~~~~~~~~~~-~ \left( \frac{245}{4} 
+ \frac{71}{2} \alpha - \frac{7}{4} \alpha^2 \right) \zeta(3) \right) C_A^2
+ 3 C_F^2 \right. \nonumber \\
&& \left. ~~~~~~~~+\, \left( 24 \zeta(3) - \frac{242}{3} - 3 \alpha^2 
- \alpha^3 \right) C_F C_A \right] a^3 ~+~ O(a^4) 
\end{eqnarray}  
which both agree in the Landau gauge and the variables in each expression 
correspond to those of the scheme indicated in the left hand side.  

\sect{Tensor current in the RI$^\prime$ scheme.} 

We now turn to the computation of the anomalous dimension for the flavour 
non-singlet tensor current at zero momentum insertion in the Landau gauge. This
calculation is similar to the one for the vector current though the 
decomposition of the Green's function $G^{\mu\nu}_{\sigma^{\mu\nu}}(p)$ will 
have different Lorentz structures. In particular we have
\begin{equation} 
G^{\mu\nu}_{\bar{\psi} \sigma^{\mu\nu} \psi}(p) ~=~  
\langle \psi(p) ~ [ \bar{\psi} \sigma^{\mu\nu} \psi ] (0) ~ \bar{\psi}(-p) 
\rangle ~=~ \Sigma^{(1)}_{\bar{\psi} \sigma^{\mu\nu} \psi}(p) 
\sigma^{\mu\nu} ~+~ \Sigma^{(2)}_{\bar{\psi} \sigma^{\mu\nu} \psi}(p) \left(
\pslash \gamma^\mu p^ \nu ~-~ \pslash \gamma^\nu p^\mu \right) \frac{1}{p^2}  
\label{tensdec} 
\end{equation}  
since $\sigma^{\mu\nu}$ is antisymmetric in its Lorentz indices. The components
are deduced from  
\begin{eqnarray} 
\Sigma^{(1)}_{\bar{\psi} \sigma^{\mu\nu} \psi}(p) &=& -~ \frac{1}{4(d-1)(d-2)} 
\left[ \mbox{tr} \left( \sigma_{\mu\nu} G^{\mu\nu}_{\bar{\psi} \sigma^{\mu\nu} 
\psi}(p) \right) ~+~ \frac{1}{p^2} \mbox{tr} \left( ( \pslash \gamma_\mu p_\nu 
- \pslash \gamma_\nu p_\mu ) G^{\mu\nu}_{\bar{\psi} \sigma^{\mu\nu} \psi}(p) 
\right) \right] \nonumber \\
\Sigma^{(2)}_{\bar{\psi} \sigma^{\mu\nu} \psi}(p) &=& -~ \frac{1}{4(d-1)(d-2)} 
\! \left[ \mbox{tr} \left( \sigma_{\mu\nu} G^{\mu\nu}_{\bar{\psi} 
\sigma^{\mu\nu} \psi}(p) \right) \,+\, \frac{d}{2p^2} \mbox{tr} \left( ( 
\pslash \gamma_\mu p_\nu - \pslash \gamma_\nu p_\mu ) G^{\mu\nu}_{\bar{\psi} 
\sigma^{\mu\nu} \psi}(p) \right) \right] \,. \nonumber \\
\end{eqnarray} 
As the anomalous dimension of the current had been computed for arbitrary 
covariant gauge parameter in the $\MSbar$ scheme, \cite{35,36,34}, we note that
the definition of the RI$^\prime$ scheme renormalization constant is
\begin{equation}  
\left. \lim_{\epsilon \, \rightarrow \, 0} \left[ 
Z^{\mbox{\footnotesize{RI$^\prime$}}}_\psi  
Z^{\mbox{\footnotesize{RI$^\prime$}}}_{\bar{\psi} \sigma^{\mu\nu} \psi}  
\Sigma^{(1)}_{\bar{\psi} \sigma^{\mu\nu} \psi}(p) \right] \right|_{p^2 \, = \, 
\mu^2} ~=~ 1 ~. 
\end{equation}  
Therefore, we find that the gauge dependent renormalization constant is 
\begin{eqnarray} 
Z^{\mbox{\footnotesize{RI$^\prime$}}}_{\bar{\psi} \sigma^{\mu\nu} \psi} &=& 
1 ~+~ \left ( \frac{C_F}{\epsilon} + \alpha C_F \right) a ~+~ \left[ 
\left( \frac{1}{2} C_F^2 - \frac{11}{6} C_F C_A + \frac{2}{3} C_F T_F \Nf 
\right) \frac{1}{\epsilon^2} \right. \nonumber \\ 
&& \left. ~+~ \left( \frac{257}{36} C_F C_A - \frac{13}{9} C_F T_F \Nf
- \left( \frac{19}{4} - \alpha \right) C_F^2 \right) \frac{1}{\epsilon} 
\right. \nonumber \\
&& \left. ~+~ \left( \left( \frac{5987}{216} + \frac{223\alpha}{36} 
+ \frac{5\alpha^2}{4} + \frac{\alpha^3}{4} - 14 \zeta(3) \right) C_F C_A 
- \left( \frac{313}{54} + \frac{20\alpha}{9} \right) C_F T_F \Nf \right. 
\right. \nonumber \\
&& \left. \left. ~~~~~~~~+~ \left( \alpha^2 - \frac{535}{24} 
+ 20 \zeta(3) \right) C_F^2 \right) \right] a^2 \nonumber \\ 
&& ~+~ \left[ \left( \frac{2}{3} C_F^2 T_F \Nf - \frac{88}{27} C_F C_A
T_F \Nf + \frac{16}{27} C_F T_F^2 \Nf^2 \right. \right. \nonumber \\
&& \left. \left. ~~~~~~+~ \frac{121}{27} C_F C_A^2 - \frac{11}{6} C_A C_F^2 
+ \frac{1}{6} C_F^3 \right) \frac{1}{\epsilon^3} \right. \nonumber \\
&& \left. ~~~~~~+~ \left( \frac{980}{81} C_F C_A T_F \Nf - \frac{104}{81} 
C_F T_F^2 \Nf^2 + \left( \frac{2\alpha}{3} - \frac{13}{3} \right) C_F^2 T_F \Nf 
- \frac{3439}{162} C_A^2 C_F \right. \right. \nonumber \\
&& \left. \left. ~~~~~~+~ \left( \frac{75}{4} - \frac{11\alpha}{6} \right) 
C_A C_F^2 + \left( \frac{\alpha}{2} - \frac{19}{4} \right) C_F^3 \right) 
\frac{1}{\epsilon^2} \right. \nonumber \\ 
&& \left. ~~~~~~+~ \left( \left( \frac{16}{3} \zeta(3) - \frac{13}{6} 
- \frac{11\alpha}{3} \right) C_F^2 T_F \Nf - \left( \frac{16}{3} \zeta(3) 
+ \frac{1004}{81} \right) C_F C_A T_F \Nf \right. \right. \nonumber \\
&& \left. \left. ~~~~~~~~~~~-~ \frac{4}{9} C_F T_F^2 \Nf^2 + \left( 
\frac{13639}{324} - \frac{40}{3} \zeta(3) \right) C_A^2 C_F \right. \right. 
\nonumber \\
&& \left. \left. ~~~~~~~~~~~+~ \left( \frac{70}{3} \zeta(3) - \frac{851}{24} 
+ \frac{40\alpha}{3} + \frac{5\alpha^2}{4} + \frac{\alpha^3}{4} \right) 
C_A C_F^2 \right. \right. \nonumber \\
&& \left. \left. ~~~~~~~~~~~+~ \left( \alpha^2 - \frac{19\alpha^2}{4} 
- \frac{145}{72} - \frac{4}{3} \zeta(3) \right) C_F^3 \right) 
\frac{1}{\epsilon} \right. \nonumber \\  
&& \left. ~~~~~~-~ \left( \left( \frac{186527}{729} + \frac{3976\alpha}{81}
+ \frac{50\alpha^2}{9} + \frac{10\alpha^3}{9} \right. \right. \right. 
\nonumber \\
&& \left. \left. \left. ~~~~~~~~~~~~~~+~ 4 \alpha \zeta(3) - \frac{2200}{27} 
\zeta(3) + 8 \zeta(4) \right) C_A C_F T_F \Nf \right. \right.  \nonumber \\
&& \left. \left. ~~~~~~~~~~~+~ \left( \frac{40\alpha^2}{9} 
+ \frac{1267\alpha}{54} - \frac{10849}{81} + 96 \zeta(3) - \frac{40\alpha}{3} 
\zeta(3) - 8 \zeta(4) \right) C_F^2 T_F \Nf \right. \right. \nonumber \\ 
&& \left. \left. ~~~~~~~~~~~-~ \left( \frac{13754}{729} + \frac{400\alpha}{81}
+ \frac{32}{27} \zeta(3) \right) C_F T_F^2 \Nf^2 \right. \right. \nonumber \\
&& \left. \left. ~~~~~~~~~~~+~ \left( \left( \frac{72491}{216} 
+ \frac{203\alpha}{12} - \frac{7\alpha^2}{8} \right) \zeta(3) 
+ \frac{44}{3} \zeta(4) + \left( \frac{10\alpha}{3} - 30 \right) \zeta(5) 
\right. \right. \right. \nonumber \\
&& \left. \left. \left. ~~~~~~~~~~~~~~~~~-~ \frac{6883865}{11664} 
- \frac{28621\alpha}{324} - \frac{157\alpha^2}{9} - \frac{421\alpha^3}{72} 
- \frac{17\alpha^4}{16} - \frac{\alpha^5}{8} \right) C_A^2 C_F \right. \right.
\nonumber \\
&& \left. \left. ~~~~~~~~~~~+~ \left( \frac{495847}{648} 
- \frac{4673\alpha}{216} - \frac{89\alpha^2}{2} - 2 \alpha^3 
- \frac{\alpha^4}{2} \right. \right. \right. \nonumber \\
&& \left. \left. \left. ~~~~~~~~~~~~~~~~~-~ \left( 530 - \frac{5\alpha}{3} 
+ \alpha^2 \right) \zeta(3) - 40 \zeta(4) - 20 \zeta(5) \right) C_A C_F^2 
\right. \right. \nonumber \\
&& \left. \left. ~~~~~~~~~~~+~ \left( \left( \frac{742}{9} \zeta(3) - 18 \alpha
+ 2 \alpha^2 \right) \zeta(3) + \frac{64}{3} \zeta(4) + 40 \zeta(5) 
- \frac{17303}{108} \right. \right. \right. \nonumber \\
&& \left. \left. \left. ~~~~~~~~~~~~~~~~~+~ \frac{523\alpha}{24} - 2 \alpha^2 
- \alpha^3 \right) C_F^3 \right) \right] a^3 ~+~ O(a^4) 
\label{ztensri} 
\end{eqnarray} 
where we have used the same symbolic manipulation programme to compute this as 
for the scalar and vector cases aside from changing the Feynman rule for the 
operator insertion. Moreover, given that the programme correctly reproduces the
gauge independent $\MSbar$ renormalization constant for all $\alpha$, 
\cite{35,36,34}, we are confident that (\ref{ztensri}) is correct. Therefore, 
from 
\begin{equation}
\gamma^{\mbox{\footnotesize{RI$^\prime$}}}_{\bar{\psi}\sigma^{\mu\nu}\psi} 
(a) ~=~ -~ \beta(a) \frac{\partial \ln 
Z^{\mbox{\footnotesize{RI$^\prime$}}}_{\bar{\psi}\sigma^{\mu\nu}\psi}} 
{\partial a} ~-~ \alpha \gamma^{\mbox{\footnotesize{RI$^\prime$}}}_\alpha(a) 
\frac{\partial \ln 
Z^{\mbox{\footnotesize{RI$^\prime$}}}_{\bar{\psi}\sigma^{\mu\nu}\psi}} 
{\partial \alpha} 
\end{equation}
we find that, in four dimensions,  
\begin{eqnarray} 
\gamma^{\mbox{\footnotesize{RI$^\prime$}}}_{\bar{\psi}\sigma^{\mu\nu}\psi}(a) 
&=& C_F a ~+~ [ (257 + 27 \alpha + 9 \alpha^2)C_A ~-~ 171C_F ~-~ 52 T_F \Nf ] 
\frac{C_F a^2}{18} \nonumber \\ 
&& +~ \left[ ( 213548 + 16902 \alpha + 5715 \alpha^2 + 1215 \alpha^3 
+ 162 \alpha^4 - 92448 \zeta(3) ) C_A^2 \right. \nonumber \\
&& \left. ~~~~-~ ( 228744 - 972 \alpha^2 - 324 \alpha^3 - 167616 \zeta(3) ) 
C_A C_F \right. \nonumber \\
&& \left. ~~~~-~ ( 99536 + 6264 \alpha + 1440 \alpha^2 - 13824 \zeta(3) ) 
C_A T_F \Nf \right. \nonumber \\ 
&& \left. ~~~~+~ ( 45576 - 24192 \zeta(3) ) C_F T_F \Nf \right. \nonumber \\ 
&& \left. ~~~~+~ ( 39420 - 41472 \zeta(3) ) C_F^2 ~+~ 9152 T_F^2 \Nf^2 \right] \frac{C_F a^3}{648} ~+~ O(a^4)  
\end{eqnarray}  
which is one of the main results of this article. For completeness, we note 
that\footnote{In \cite{34} the Casimir of the final coefficient should have 
been $T_F^2 \Nf^2$ and not $T_F^2 C_F^2$.}, \cite{35,36,34}, 
\begin{eqnarray} 
\gamma^{\mbox{\footnotesize{$\MSbar$}}}_{\bar{\psi}\sigma^{\mu\nu}\psi}(a) &=& 
C_F a ~+~ [ 257C_A ~-~ 171C_F ~-~ 52 T_F \Nf ] 
\frac{C_F a^2}{18} \nonumber \\ 
&& +~ \left[ 13639 C_A^2 ~-~ 4320\zeta(3) C_A^2 ~+~ 
12096\zeta(3)C_A C_F \right. \nonumber \\ 
&& \left. ~~~~-~ 20469 C_A C_F ~-~ 1728\zeta(3) C_A T_F \Nf ~-~ 
4016 C_A T_F \Nf \right. \nonumber \\ 
&& \left. ~~~~-~ 6912 \zeta(3) C_F^2 ~+~ 6570 C_F^2 ~+~ 
1728\zeta(3) C_F T_F \Nf \right. \nonumber \\ 
&& \left. ~~~~+~ 1176 C_F T_F \Nf ~-~ 144 T_F^2 \Nf^2) \right] 
\frac{C_F a^3}{108} ~+~ O(a^4) 
\label{siganom} 
\end{eqnarray}  
where the four loop expression of (\ref{siganom}) is available for QED in the
quenched approximation, \cite{36}. Specifying the Landau gauge for the colour 
group $SU(3)$ we find
\begin{eqnarray} 
\left. \gamma^{\mbox{\footnotesize{RI$^\prime$}}}_{\bar{\psi}\sigma^{\mu\nu}
\psi}(a) \right|^{SU(3)}_{\alpha \, = \, 0} &=& \frac{4}{3} a ~-~ 
\frac{2}{27} [ 26 \Nf - 543 ] a^2 \nonumber \\ 
&& \! +\, \frac{2}{243} \! \left[ 572 \Nf^2 + ( 1152 \zeta(3) - 29730 ) \Nf 
- 58824 \zeta(3) + 269259 \right] \! a^3 ~+~ O(a^4) \nonumber \\  
\end{eqnarray} 
where we have set $T_f$ $=$ $1/2$, $C_F$ $=$ $4/3$ and $C_A$ $=$ $3$. Finally,
we note  
\begin{eqnarray} 
\left. \Sigma^{(1) ~ \mbox{\footnotesize{$\MSbar$}} ~
\mbox{\footnotesize{finite}}}_{\bar{\psi} \sigma^{\mu\nu} \psi}(p) 
\right|_{p^2 \, = \, \mu^2} &=& 1 ~-~ \left[ \left( 
\frac{3773}{216} - 3 \alpha - \frac{3}{8} \alpha^2 - 11 \zeta(3) + 3 \zeta(3) 
\alpha \right) C_F C_A \right. \nonumber \\
&& \left. ~~~~~~~~~~-~ \frac{62}{27} T_F \Nf C_F + \left(
\alpha^2 - \frac{65}{3} + 20 \zeta(3) \right) C_F^2 \right] a^2 \nonumber \\
&& -~ \left[ \left( \frac{23831}{162} - \frac{4}{3} \alpha - 112 \zeta(3)
- \frac{8}{3} \zeta(3) \alpha + 8 \zeta(4) \right) T_F \Nf C_F^2 \right. 
\nonumber \\
&& \left. ~~~~~+~ \left( \frac{673}{72} \alpha - \frac{79544}{729} 
+ \frac{1732}{27} \zeta(3) \right. \right. \nonumber \\ 
&& \left. \left. ~~~~~~~~~~~~-~ 4 \zeta(3) \alpha - 8 \zeta(4) \right) T_F \Nf 
C_A C_F \right. \nonumber \\
&& \left. ~~~~~+~ \left( \frac{32}{27} \zeta(3) - \frac{376}{729} \right) 
T_F^2 \Nf^2 C_F \right. \nonumber \\
&& \left. ~~~~~+~ \left( \frac{4017239}{11664} - \frac{12817}{576} \alpha 
- \frac{197}{64} \alpha^2 - \frac{29}{48} \alpha^3 \right. \right. \nonumber \\
&& \left. \left. ~~~~~~~~~~~~-~ \left( \frac{5530}{27} - \frac{253}{12}
\alpha - \frac{15}{4} \alpha^2 - \frac{1}{3} \alpha^3 \right) \zeta(3) 
\right. \right. \nonumber \\
&& \left. \left. ~~~~~~~~~~~~-~ \left( \frac{497}{48} + \frac{3}{8} \alpha 
+ \frac{3}{16} \alpha^2 \right) \zeta(4) \right. \right. \nonumber \\
&& \left. \left. ~~~~~~~~~~~~-~ \left( \frac{45}{4} + \frac{35}{6} \alpha 
+ \frac{5}{4} \alpha^2 \right) \zeta(5) \right) C_A^2 C_F \right. \nonumber \\
&& \left. ~~~~~+~ \left( \left( 486 + \frac{79}{3} \alpha - 2 \alpha^2 
- \alpha^3 \right) \zeta(3) + 34 \zeta(4) \right. \right. \nonumber \\
&& \left. \left. ~~~~~~~~~~~~+~ \left( 40 - 20 \alpha \right) \zeta(5) 
- \frac{58616}{81} \right. \right. \nonumber \\
&& \left. \left. ~~~~~~~~~~~~+~ \frac{1}{6} \alpha + 6 \alpha^2 + \frac{3}{2} 
\alpha^3 \right) C_F^2 C_A \right. \nonumber \\
&& \left. ~~~~~+~ \left( \frac{4490}{27} - \alpha + 2 \alpha^2 
- \frac{64}{3} \zeta(4) - 40 \zeta(5) \right. \right. \nonumber \\
&& \left. \left. ~~~~~~~~~~~~-~ \left( \frac{742}{9} + 2 \alpha + 2 \alpha^2 
- \frac{2}{3} \alpha^3 \right) \zeta(3) \right) C_F^3 \right] a^3 \nonumber \\
&& +~ O(a^4) 
\end{eqnarray} 
which implies 
\begin{eqnarray} 
\left. \Sigma^{(1) ~ \mbox{\footnotesize{$\MSbar$}} ~
\mbox{\footnotesize{finite}}}_{\bar{\psi} \sigma^{\mu\nu} \psi}(p) 
\right|^{SU(3) \, , \, \alpha \, = \, 0}_{p^2 \, = \, \mu^2} &=& 1 ~-~ 
\left( \frac{1693}{54} - \frac{76}{9} \zeta(3) - \frac{124}{81} \Nf \right) 
a^2 \nonumber \\
&& -~ \left( \frac{1946885}{2916} - \frac{14872}{243} \zeta(3)
+ \frac{2111}{324} \zeta(4) - \frac{445}{27} \zeta(5) \right. \nonumber \\
&& \left. ~~~~~~~+~ \left( \frac{776}{27} \zeta(3) - \frac{80}{9} \zeta(4)
- \frac{63764}{729} \right) \Nf \right. \nonumber \\
&& \left. ~~~~~~~+~ \left( \frac{32}{81} \zeta(3) - \frac{376}{2187} \right) 
\Nf^2 \right) a^3 ~+~ O(a^4) ~.  
\end{eqnarray} 
For $\left. \Sigma^{(2) ~ \mbox{\footnotesize{$\MSbar$}} ~ 
\mbox{\footnotesize{finite}}}_{\bar{\psi} 
\sigma^{\mu\nu} \psi}(p) \right|_{p^2 \, = \, \mu^2}$ and $\left. \Sigma^{(2) ~
\mbox{\footnotesize{RI$^\prime$}} ~ 
\mbox{\footnotesize{finite}}}_{\bar{\psi} \sigma^{\mu\nu} \psi}(p) 
\right|_{p^2 \, = \, \mu^2}$ it transpires that there are no contributions to 
the finite part to and including three loops. This is consistent with the one 
loop evaluation of the same quantity in \cite{37} when the chiral limit is 
taken. For completeness we have again checked our RI$^\prime$ scheme anomalous 
dimension by constructing the conversion function 
$C_{\bar{\psi} \sigma^{\mu\nu} \psi}(a,\alpha)$ explicitly. From   
\begin{eqnarray} 
C_{\bar{\psi} \sigma^{\mu\nu} \psi}(a,\alpha) &=& 1 ~+~ \alpha C_F a ~+~  
\left[ \left( 216 \alpha^2 + 4320 \zeta(3) - 4815 \right) C_F - 1252 T_F \Nf 
\right. \nonumber \\
&& \left. ~~~~~~~~~~~~~~~~~~~~~+~ \left( 162 \alpha^2 + 756 \alpha 
- 3024 \zeta(3) + 5987 \right) C_A \right] \frac{C_F a^2}{216} \nonumber \\
&& +~ \left[ \left( 23328 \alpha^3 + 104976 \alpha^2 + 23328 \alpha^2 \zeta(3) 
+ 504684 \alpha - 38880 \alpha \zeta(3) \right. \right. \nonumber \\
&& \left. \left. ~~~~~+~ 12363840 \zeta(3) + 933120 \zeta(4) + 466560 \zeta(5) 
- 17850492 \right) C_A C_F \right. \nonumber \\
&& \left. ~~~~+~ \left( 39366 \alpha^3 - 26244 \alpha^2 \zeta(3)
+ 215055 \alpha^2 - 324648 \alpha \zeta(3) - 77760 \alpha \zeta(5) \right. 
\right. \nonumber \\
&& \left. \left. ~~~~~~~~~~+~ 1092771 \alpha - 7829028 \zeta(3) 
- 342144 \zeta(4) \right. \right. \nonumber \\
&& \left. \left. ~~~~~~~~~~+~ 699840 \zeta(5) + 13767730 \right) C_A^2 \right. 
\nonumber \\
&& \left. ~~~~+~ \left( 93312 \alpha \zeta(3) - 340200 \alpha 
+ 1900800 \zeta(3) \right. \right. \nonumber \\
&& \left. \left. ~~~~~~~~~~-~ 186624 \zeta(4) - 5968864 \right) C_A T_F \Nf 
\right. \nonumber \\
&& \left. ~~~~-~ \left( 62208 \alpha \zeta(3) + 119664 \alpha 
+ 2239488 \zeta(3) \right. \right. \nonumber \\
&& \left. \left. ~~~~~~~~~~-~ 186624 \zeta(4) - 3124512 \right) C_F T_F \Nf 
\right. \nonumber \\
&& \left. ~~~~+~ \left( 23328 \alpha^3 - 46656 \alpha^2 \zeta(3) 
+ 46656 \alpha^2 + 419904 \alpha \zeta(3) - 508356 \alpha \right. \right. 
\nonumber \\
&& \left. \left. ~~~~~~~~~~-~ 1923264 \zeta(3) - 497664 \zeta(4) 
- 933120 \zeta(5) + 3737448 \right) C_F^2 \right. \nonumber \\
&& \left. ~~~~+~ \left( 27648 \zeta(3) + 440128 \right) T_F^2 \Nf^2 \right] 
\frac{C_F a^3}{23328} ~+~ O(a^4) 
\end{eqnarray} 
we again have exact agreement. 

\sect{Discussion.} 

To conclude we have first renormalized QCD to three loops in arbitrary 
covariant gauge in the RI$^\prime$ scheme. The full renormalization was 
necessary since, for example, the anomalous dimension of the gauge parameter is
required when converting the renormalization constants to renormalization group functions for non-zero $\alpha$. Although in practice one only requires
information in the Landau gauge, computing for $\alpha$~$\neq$~$0$ provides
important internal checks on the calculation such as for comparing with 
established $\MSbar$ results. Further, we have extended the machinery of 
\cite{12} to compute the anomalous dimensions of the tensor operator 
$\bar{\psi} \sigma^{\mu\nu} \psi$ at three loops in the chiral limit in the 
Landau gauge in a scheme which is natural in lattice regularization. Given this
approach it would be interesting to examine other operators whose anomalous 
dimensions are required in the RI$^\prime$ scheme such as the low moments of 
the twist-$2$ Wilson operators which occur in the operator product expansion in
deep inelastic scattering and those relating to transversity, in order to 
provide the foundation to improve lattice estimates of matrix elements. 

\vspace{1cm} 
\noindent
{\bf Acknowledgements.} The author thanks Dr P.E.L. Rakow and Dr C. McNeile for
valuable discussions.

\end{document}